\documentclass[11pt,twoside,a4paper]{article}
\usepackage[affil-it]{authblk}
\usepackage{ amssymb }
\usepackage{amsmath}
\usepackage{tikz}
\usepackage{setspace}
\usepackage{tensor}
\usetikzlibrary{matrix}
\usepackage[margin=2.9 cm]{geometry}
\usepackage[cp1250]{inputenc}
\usepackage{polski}
\usepackage[polish,english]{babel}
\usepackage[toc,page]{appendix}

\usepackage[colorlinks,citecolor=green,urlcolor=blue,bookmarks=false,hypertexnames=true]{hyperref}

\input xy
\xyoption{all} \tolerance=500

\def\sT{\mathsf T}

\def\d{\mathrm d}

\def\sV{\mathsf V}

\def\sJ{\mathsf J^1}

\def\J2{\mathsf J^2}

\def\wdt{\widetilde }
\def\bh{\textbf{h}}

\newtheorem{theorem}{Theorem}

\newtheorem{lemma}{Lemma}

\title{Cauchy data space and multisymplectic formulation of conformal classical field theories}

\begin{document}

\author{O\u{g}ul Esen$^{\dagger}$, Manuel de Le\'on$^{\ddagger}$, Cristina
Sard\'on$^{*}$,  Marcin Zaj\k{a}c$^{**}$}
\maketitle

\centerline{Department of Mathematics$^{\dagger}$,}
\centerline{Gebze Technical University, 41400 Gebze, Kocaeli, Turkey.}
\centerline{oesen@gtu.edu.tr}
\vskip 0.5cm

\centerline{Instituto de Ciencias Matem\'aticas, Campus Cantoblanco$^{\ddagger}$,}
\centerline{Consejo Superior de Investigaciones Cient\'ificas}
\centerline{and}
\centerline{
Real Academia Espa{\~n}ola de las Ciencias.}
\centerline{C/ Nicol\'as Cabrera, 13--15, 28049, Madrid, Spain.}
\centerline{mdeleon@icmat.es}
\vskip 0.5cm

\centerline{Instituto de Ciencias Matem\'aticas, Campus Cantoblanco$^{*}$,}
\centerline{Consejo Superior de Investigaciones Cient\'ificas.}
\centerline{C/ Nicol\'as Cabrera, 13--15, 28049, Madrid, Spain.}
\centerline{cristinasardon@icmat.es}
\vskip 0.5cm

\centerline{Department of Mathematical Methods in Physics$^{**}$,}
\centerline{Faculty of Physics. University of Warsaw,}
\centerline{ul. Pasteura 5, 02-093 Warsaw, Poland.}
\centerline{
marcin.zajac@fuw.edu.pl}

\begin{abstract}

In this article we inspect the dynamics of classical field theories
with a local conformal behavior. Our interest in the multisymplectic
setting comes from its suitable description of field theories, and the conformal character has been added to account for field theories that are scale invariant, flat spaces, and because some conformal fields can be exactly solved or classified. In particular, we will solve the example of a conformal scalar field using the geometric Hamilton--Jacobi theory that is explicitly proposed for conformal fields on a multisymplectic manifold. To complete the geometric approach to study field theories, we propose the Hamilton--Jacobi theory for conformal fields in a Cauchy data space, in which space and time are split separately and the dynamics is depicted in an infinite-dimensional manifold.
\end{abstract}

\tableofcontents

\setlength{\parindent}{0em}
\setlength{\parskip}{0.5cm}

\section{Introduction}
\onehalfspacing
The construction of field theories in Physics has a long history, going back to Volterra's work in the XIX century, when a classical field theory was pursued to predict how one or more physical fields interact with matter through field equations. From there, the term classical 
field theory was coined for physical theories as electromagnetism and gravitation, and
also incorporated to quantum mechanics as well. 

Classical field theory deals with fields, or quantities asigned at each point of space 
and time, like the value of an electromagnetic field at a particular point. Hence, the
classical Hamiltonian framework of classical mechanics needed to be extended to these geometric objects called fields. Analogously to the case of mechanics, we propose the study of general multi-phase spaces, where the setting of a symplectic manifold is replaced by a multisymplectic one, which is not only important to develop a field theory, but also for the quantization of mechanics. The geometric foundations of field theory and variational calculus were not established propertly until the 60's, with the advent of gauge theories \cite{AA,KT,LeonRodri,LeonSalVil}. We cite \cite{GoIsMaMo98,GoIsMa04} for some fundamentals.

Inspired by conformal symplectic structures as a generalization of symplectic
structures, we have generalized the concept of multisymplectic structure to conformal multisymplectic structures, which are structures that locally are equivalent to a multisymplectic manifold, but the local multisymplectic structure is only well-defined up to scaling
by a constant, and the monodromy of the local multisymplectic structure around curves
may induce these rescalings \cite{vaisman}.

So, one main aim of this paper is to generalize the locally conformal symplectic formulation of nonautonomous mechanics to first-order locally conformal multisymplectic field theories. We aim at pursuing field theories whose Hamiltonians or dynamical variables are defined in open subsets of a manifold. In those subsets, these systems behave geometrically like multisymplectic fields, although the complete global dynamics fails to be multisymplectic. This phenomenom appears,
for example, in some field theories with nonlocal potentials. Field theories with such local and global characteristics will be referred to as dynamical systems on a locally
conformal multisymplectic (l.c.m.-s) manifolds, from a geometric point of view. Let us explain this local
setting more explicitly.

One can exploit the parallelism between Classical Mechanics and Classical
field theories, but one needs to be very careful. Indeed, instead of a configuration manifold, we have now a configuration bundle $\pi:E\rightarrow M$ such that its sections are the fields and the $m$ dimensional manifold $M$ is the space-time manifold.

The Lagrangian density depends on the space coordinates (and time), the fields and their derivatives, so it is very natural to take the manifold of 1-jets of sections of $\pi$, i.e., $J^1\pi$, as the generalization of the tangent bundle in Classical Mechanics.
Then, a Lagrangian density is a fibered mapping $\mathcal{L}:J^1\pi\rightarrow \Lambda^m E$. From the Lagrangian density one can construct
the Poincaré-Cartan form which gives the evolution of the system.
On the other hand, the spaces of 1- and 2-horizontal m-forms on $E$ with respect to the projection $\pi$, denoted respectively by $\Lambda^m_2E$ and $\Lambda^m_1E$, are the bundles where the Hamiltonian picture of the theory is developed. To be more
precise, the phase space is just the quotient $J^1\pi^*=\Lambda^m_2E/\Lambda^m_1E$. The Hamiltonian density is a section of $\Lambda^m_2E\rightarrow J^1\pi^*$ (the Hamiltonian
function appears when a volume form $\eta$ on $M$ is chosen, such that $\mathcal{H}=H\eta$. The
Hamiltonian section $\mathcal{H}$ permits just to pull-back the canonical multisymplectic of $\Lambda^m_2E$ to a multisymplectic form on $J^1\pi^*$.
Of course, both descriptions are related by the Legendre transform which
send solutions of the Euler-Lagrange equations into solutions of the Hamilton
equations. One important difference with the case of mechanics is that now
we are dealing with partial differential equations and we lost in principle the
integrability. In any case, the solutions in the Lagrangian, as well as in the Hamiltonian side can be interpreted as integral
sections of Ehresmann connections.

Another motivation for this formulation is the following. As is well known, there is no canonical model for the Hamiltonian first-order field theory. To overcome difficulties in the definition of Poisson brackets, reduction, etc., we use the extended multi-momentum bundle, denoted by $\Lambda^m_2E$, which is analogous to the extended phase space $T^{*}(Q\times \mathbb{R})$ of a nonautonomous mechanical system. $\Lambda^m_2E$ has a canonical multisymplectic form since it is a vector subbundle of a multicotangent bundle. In this manifold  the physical information is given by a Hamiltonian (one-)form. Then, Hamiltonian systems can be introduced as in autonomous mechanics, by using certain kinds of Hamiltonian multivector fields. The resultant extended Hamiltonian formalism is the generalization to field theories of the extended formalism for nonautonomous mechanical systems \cite{EcheLeon,Paufler}.


There exists an alternative way of studying classical
field theories, this one involves an infinite dimensional setting. The idea is to split the spacetime manifold $E$ in time and space. To do this, we need to take a Cauchy surface, that is, a $n$-dimensional submanifold $N$ of $M$ such that (at
least locally) we have $M=\mathbb{R}\times N$. So, the space of embeddings from $N$ to $J^1\pi^*$
is known as the Cauchy space of data for a particular choice of a Cauchy surface.
This allows us to integrate the multisymplectic form on $J^1\pi^*$ to the Cauchy data
space and obtain a presymplectic (indeed, precosymplectic) infinite dimensional system, whose dynamics
is related to the HDW equations for $H$ \cite{CaLeDiVa15}.

In order to reproduce the dynamics in conformal multisymplectic manifolds and in the conformal 
Cauchy data space, we shall propose a Hamilton--Jacobi theory (HJ theory). There are several choices for describing the dynamics of dynamical systems and physical fields, from a Newtonian to a Hamiltonian formalism. Our choice is the Hamilton--Jacobi theory, it probably constitutes the third
most important theory in classical mechanics and it is very useful under certain circumstances. For example, when the forces acting on the physical problem derive from a separable potential, the Hamilton--Jacobi equation can be our preferred option. Let us recall such equation. In the case of the time-independent Hamilton-Jacobi equation (HJE) takes the form:
\begin{equation}\label{HJeq1}
H\left(q^i,\frac{\partial W}{\partial q^i}\right)=E
\end{equation}
with $E$ is a constant.
In the symplectic case, this equation can be interpreted geometrically with a primordial observation that if a Hamiltonian
vector field $X_{H}:T^{*}Q\rightarrow TT^{*}Q$ can be projected into a vector field $X_H^{dW}:Q\rightarrow TQ$ on a lower
dimensional manifold by means of a 1-form $dW$, then the integral curves of the projected
vector field $X_{H}^{dW}$ can be transformed into integral curves of $X_{H}$ provided that $W$ is a solution of (\ref{HJeq1}). If we define the projected vector field as:
\begin{equation}
 X_H^{dW}=T\pi\circ X_H\circ dW,
\end{equation}
where $T{\pi}$ is the induced projection on the tangent space, $T{\pi}:TT^{*}Q\rightarrow T^{*}Q$ by the canonical projection
$\pi:T^{*}Q\rightarrow Q$, it implies the commutativity of the diagram below:

\[
\xymatrix{ T^{*}Q
\ar[dd]^{\pi} \ar[rrr]^{X_H}&   & &TT^{*}Q\ar[dd]^{T\pi}\\
  &  & &\\
 Q\ar@/^2pc/[uu]^{dW}\ar[rrr]^{X_H^{dW}}&  & & TQ}
\]
\noindent
where Im$dW$ is a Lagrangian submanifold, since $dW$ is exact and then, it is closed. This condition gave rise to the introduction of Lagrangian submanifolds, which are very important objects in Hamiltonian mechanics, since the dynamical equation (Hamiltonian or Lagrangian)
can be described as a Lagrangian submanifolds of convenient symplectic manifolds. So, we enunciate the following theorem.

\begin{theorem}
\label{HJT} For a closed one-form $\gamma=dW$ on $Q$ the following conditions are
equivalent:
\begin{enumerate}
\item The vector fields $X_{H}$ and $X_{H}^{\gamma }$ are $\gamma$-related,
that is
\begin{equation}
T\gamma \circ X_H^{\gamma}=X_H\circ\gamma.
\end{equation}
\item The following equation is fulfilled 
\[
d\left( H\circ \gamma \right)=0.
\]
\end{enumerate}
\end{theorem}

\noindent The first item in the theorem says that if $\left( q^i\left(
t\right) \right) $ is an integral curve of $X_{H}^{\gamma }$, then $\left(
q^i\left( t\right) ,\gamma_j \left( q\left( t\right) \right) \right) $ is an
integral curve of the Hamiltonian vector field $X_{H}$, hence a solution of
the Hamilton equations. The solution of the Hamiltonian
equations is called horizontal since it is on the image of a one-form on $Q.$ In the local picture, the second condition implies that exterior
derivative of the Hamiltonian function on the image of $\gamma $ is closed,
that is, $H\circ \gamma $ is constant. Under the assumption that $\gamma $ is closed, we can find (at least
locally) a function $W$ on $Q$ satisfying $dW=\gamma $. Here $\gamma$ is the solution of the HJ equation and it is a Lagrangian submanifold on $T^{*}Q$. It was Tulczyjew who pioneered the characterization of the image of local Hamiltonian vector fields on a symplectic manifold $(M,\omega)$ as Lagrangian submanifolds
of a symplectic manifold $(TM,\omega^{T})$, where $\omega$ is the tangent lift of $\omega$ to $TM$ \cite{Tulczy}. This result was later generalized to Poisson manifolds \cite{GrabUrb} and Jacobi manifolds \cite{IbaLeonMarrDiego1}.
The HJ theory and the corresponding Lagrangian submanifolds have also been proposed in multisymplectic settings \cite{LeonPrietoRRVil}
and here we aim to extend it to conformal multisymplectic field theories.

The outline of the paper goes as follows.  In Section 2 we review the fundamentals of multisymplectic manifolds and Hamiltonian dynamics on such manifolds. We recall the Hamilton-De Donder-Weyl equations and Ehresmann connection and propose a strictly geometrical Hamilton--Jacobi theory on multisymplectic manifolds, in comparison to the former Hamilton--Jacobi theorems formulated in coordinates. In Section 3 we introduce the notion of locally conformal multisymplectic framework and the fundamentals of geometry on such manifolds. We propose the HDW equations on locally conformal multisymplectic manifolds and a Hamilton-Jacobi theory in terms of a connection.
 Section 4 contains two examples: the problem of conformal time-dependent mechanics and conformal scalar fields. In Section 5 we go a step forward with our theory by considering a infinite dimensional space, i.e., the dynamics will be described in a Cauchy data space. As a last result, we present a Hamilton-Jacobi theory for conformal fields in the Cauchy data space.
 
From now on, assume all mathematical objects to be $C^{\infty}$, globally defined and that manifolds are connected. This permits us to suit technical
details while highlightning the main aspects of the theory.

\section{Hamiltonian Mechanics in a Multisymplectic Framework}

\subsection{Multisymplectic Manifolds} \label{Sec-MM}

A closed and $1$-nondegenerate $r$-form $\Omega$ on a manifold $P$ is called a multisymplectic $r$-form, and the pair $(P,\Omega)$ is called an ($r$-)multisymplectic manifold, see for example \cite{AA,BSF88,CIdL,CIdL2,Go91}. Recall that a $r$-form is $1$-nondegenerate if the equation $\iota_V\Omega=0$ implies $V=0$, with $V\in TP$. 

\textbf{Submanifolds.} Consider a vector subbundle $F$ of the tangent bundle $TP$ of a multisymplectic manifold $P$. The $l$-orthogonal complement $F^{\perp,l}$ of $F$ is also a vector subbundle of $TP$ defined pointwisely by 
\begin{equation} \label{l-ortho}
F^{\perp,l}_p=\{V_p\in T_pP : \Omega_p(V_p,V_1,\dots, V_l)=0, \quad \forall V_1, \dots, V_l \in F_p\}.
\end{equation} 
Let $S$ be a submanifold of a multisymplectic manifold $P$. $S$ is said to be $l$-isotropic submanifold if the tangent space $T_pS$ is a subspace of its $l$-orthogonal $T_pS^{\perp,l}$, we say that $S$ is a $l$-coisotropic submanifold  if $T_pS^{\perp,l}$ is a subspace of $ T_pS$, and it is said that $S$ is a $l$-Lagrangian submanifold  if $T_pS$ is precisely equal to $T_pS^{\perp,l}$ for all $p$ in $S$. We particularly call $(r-1)$-Lagrangian submanifold of an $r$-multisymplectic manifold as Lagrangian submanifold.   

\textbf{Bundle of k-forms.}
Let $N$ be a smooth manifold, and $\Lambda^kN$ be the bundle of $k$-forms on $N$. With the projection $\zeta$ from $\Lambda^kN$ to $N$, the three-tuple $(\Lambda^kN,\zeta,N)$ turns out to be a fiber bundle. There is a tautological $k$-form $\Theta$ over the total space $\Lambda^kN$. For any $\omega$ in $\Lambda^kN$, the value of $\Theta(\omega)$ over $k$ vectors, denoted by $v_1,...,v_k$, is given by
\begin{equation} \label{Theta}
\Theta(\omega)(v_1,...,v_k)=\omega(\sT_\omega\zeta(v_1),...,\sT_\omega\zeta(v_k)).
\end{equation}
Then, minus of the exterior derivative $\Omega=-\d\Theta$ becomes a multisymplectic $k+1$-form on $\Lambda^kN$. The pair $(\Lambda^kN,\Omega)$ is a generic example of multisymplectic manifolds. For an arbitrary $k$-form $\kappa$ on the base manifold $N$, that is a section $\Lambda^kN\mapsto N$, it is straight forward calculation to show that
\begin{equation} \label{alphaomega}
\kappa^*\Theta=\kappa, \qquad 
\kappa^*\Omega=-d\kappa.
\end{equation}
As a manifestation of the latter identity, it is immediate to see that, a $k$-form $\kappa$ on $N$ is closed if and only if $\kappa^*\Omega$ vanishes identically. This implies that the image space of a section $\kappa$ is a Lagrangian submanifold if and only if it is closed. 

\textbf{The space of two-horizontal forms.} Consider a fiber bundle $(E,\pi,M)$ where the base manifold $M$ is $m$-dimensional manifold equipped with a volume form $\eta$. We denote the space of $m$-forms on $E$ by $\Lambda^m E$, and define  
a vector subbundle of $\Lambda^m E$, denoted by $\Lambda^m_2E$, consisting of two-horizontal  differential forms with respect to the fibration $\pi$. In the literature, $\Lambda^m_2E$ is called multi-momentum bundle as well. Pointwisely, for any $e$ in $E$, 
\begin{equation} \label{multimom}
(\Lambda^m_2E)_e=\{\omega\in\Lambda^m_eE: \iota_{v_2}\iota_{v_1}\omega=0, \quad \forall v_1,v_2\in\sV_e\pi   \},
\end{equation}
where $\sV \pi$ is the vertical bundle with respect to the projection $\pi$. Being a bundle of $m$-forms, $\Lambda^m E$ is a multisymplectic manifold equipped with the canonical $m$-form $\Theta$ and multisymplectic $(m+1)$-form $\Omega$ as presented in the previous paragraph. We restrict these forms to the subbundle $\Lambda^m_2E$ and denote these restricted differential forms by $\Theta_2$ and $\Omega_2$, respectively. It is possible to show that the pair $(\Lambda^m_2E,\Theta_2)$ turns out to be a multisymplectic manifold \cite{BSF88}. In the light of the second identity in (\ref{alphaomega}) we can argue that the image space of the section $\bar{\gamma}$ of the fibration $\Lambda^m_2E\mapsto E $ is a Lagrangian submanifold of the multisymplectic manifold $\Lambda^m_2E$ if and only if it is closed. 

\textbf{The dual jet bundle.}  Consider now the multisymplectic manifold $(\Lambda^m_2E,\Omega_2)$ defined in the previous paragraph. Denoting the bundle of horizontal $m$-forms on $E$ as $\Lambda^{m}_1E$, we define a new bundle by taking the quotient 
\begin{equation} \label{J*}
\mathsf J^1\pi^*=\Lambda^m_2E/\Lambda^{m}_1E.
 \end{equation} 
  This quotient bundle coincides with the affine dual of the first jet bundle $J^1\pi$, i.e., it is $J^1\pi^*$ of the fiber bundle $\pi:E\mapsto M$, see \cite{RoSa14}. Further, there exists a canonical projection $\mu$ that maps a  differential form in $\Lambda^m_2E$ to the equivalence class in $\mathsf J^1\pi^*$ that it belongs to. Hence, the triple $(\Lambda^m_2E,\mu,\mathsf  J^1\pi^*)$ is a principal fiber bundle with a structure group diffeomorphic to $\mathbb R$. Here, the action of the group $\mathbb R$ on the total space $\Lambda^m_2E$ with translations is of the following form
\begin{equation} \label{R-action}
 \mathbb R\times \Lambda^m_2E\to \Lambda^m_2E, \qquad (t,\omega)\longmapsto t\cdot \eta+\omega,    
 \end{equation}
 where $\eta$ denotes the volume form on $M$ and its the pullback to any fibered bundle over $M$ will be denoted by the same letter.
Both $\Lambda^m_2E$ and $\mathsf J^1\pi^*$ are vector bundles over $E$ with projections $\bar{\nu}$, and ${\nu}$, respectively. We compose these projections with $\pi$ and arrive at two projections $\bar{\tau}:=\pi \circ \bar{\nu}$ and $\tau:=\pi \circ \nu$ from $\Lambda^m_2E$ and $\mathsf J^1\pi^*$ to the base space $M$. The following diagram illustrates the projections and the sections in one look. Notice that, $\bar{\phi}$ and $\phi$ are sections of the fibrations  $\bar{\tau}$ and $\tau$, respectively. 
 \begin{equation} \label{fibrations}
\xymatrix{ \Lambda^m_2E \ar[rr]_\mu 
\ar[dr]^{\bar{\nu}} \ar[ddr]_{\bar{\tau}} && \mathsf  J^1\pi^*   \ar[dl]_{\nu} \ar@/_1pc/[ll]_{h}  \ar[ddl]^{ \tau} 
  \\
&E \ar[d]^\pi
 \\
& M \ar@/^1pc/[uul]^{\bar{\phi}} \ar@/_1pc/[uur]_{ \phi} 
}
\end{equation}

\textbf{Local coordinates.} Assume that, $M$ is $m$-dimensional with local coordinates $(x^i)$, and $\d_mx$ is local realization of the volume form $\eta$ on $M$. On $E$, we can find bundle coordinates  $(x^i,u^\alpha)$ in which the volume form has that particular form. Then on the space of two-horizontal $m$-forms $\Lambda^m_2E$, the induced coordinates are computed to be $(x^i,u^\alpha,p,p^i_\alpha)$, so that  $\Theta_2$ and $\Omega_2$ become
\begin{equation} \label{can-forms}
\begin{split}
\Theta_2&=p\d_mx+p{^i_\alpha}\d u^\alpha\wedge(\partial_i\lrcorner\d_mx)
\\
\Omega_2&=-\d p\wedge\d_mx-\d p{^i_\alpha}\wedge\d u{^\alpha}\wedge(\partial_i\lrcorner\d_mx),
\end{split}
\end{equation}
respectively. Here, $\partial_i\lrcorner\d_mx$ denotes the contraction of the volume form $\d_mx$ with 
the vector field $\partial/\partial x^i$. In this local realization, the action in (\ref{R-action}) turns out to be 
\begin{equation}
\mathbb R\times \Lambda^m_2E \longrightarrow \Lambda^m_2E, \qquad(t,(x^i,u^\alpha, p, p^i_\alpha))\mapsto ((x^i,u^\alpha, t+p, p^i_\alpha)). 
 \end{equation}
The infinitesimal generator $Y_\mu$ of this motion is computed to be ${\partial}/{\partial p}$. This one-dimensional vector field generates the vertical bundle  $\sV\mu$ as well. So that, the induced coordinates on the quotient space $\mathsf J^1\pi^*$ are $(x^i,u^\alpha,p^i_\alpha )$. 

\subsection{Hamiltonian Dynamics in Multisymplectic Manifolds} \label{HDW-Sec}

Consider again a fiber bundle $(E,\pi,M)$, and recall the principal bundle $(\Lambda^m_2E,\mu,\mathsf  J^1\pi^*)$ presented in the previous subsection. A section, denoted by $h$,
 of the fibration $\mu$ is called a Hamiltonian section. In the multisymplectic realization of the field theory, the Hamiltonian dynamics is encoded in those sections \cite{CaLeDiVa15,E-EM-L-RR}. 

 \textbf{Hamiltonian sections and Hamiltonian densities.} We denote the infinitesimal generators of the action (\ref{R-action}) by $Y_\mu$. It can be easily established that, for a Hamiltonian section $h$, there exists a corresponding real valued function, say $\bar H$, satisfying that $Y_\mu(\bar H)$ is the unity. Further, again for a Hamiltonian section $h$, there exists a unique Hamiltonian density, that is a fiber preserving map $\mathcal H$ from $\Lambda^m_2E$ to the space $\Lambda^{m}M$ of top-forms on $M$ such that $\iota_{Y_\mu}d\mathcal H$ equals $\eta$. For a Hamiltonian section $h$, the corresponding Hamiltonian density is defined to be
\begin{equation} \label{Ham-sec-dens}
 \mathcal H: \Lambda^m_2E \longrightarrow \Lambda^{m}M, \qquad \omega\mapsto \omega-h(\mu(\omega)).
  \end{equation}
Conversely, for a Hamiltonian density $\mathcal H$, the corresponding Hamiltonian section is characterized by the condition that the image space of $h$ is the preimage of $0$ under $\mathcal H$.

\textbf{Hamilton-De Donder-Weyl (HDW) equation.} We present two equivalent definitions of the Hamilton-De Donder-Weyl (HDW) equation. The first one is defined in terms of Hamiltonian densities. Let $\mathcal H$ be a Hamiltonian density. Consider the fiber bundle $(\Lambda^m_2E,\bar{\tau},M)$ where the projection is defined to be $\bar{\tau}=\pi\circ\bar{\nu}$, see the diagram in \eqref{fibrations}. A critical point of $\mathcal H$ is a (local) section $\bar{\phi}$ of the bundle $\bar{\tau}$ that satisfies the (extended) Hamilton-De Donder-Weyl equation
\begin{equation} \label{HDW1}
\bar{\phi}^*\iota_{\bar{X}}(\Omega_2+d\mathcal H)=0,
\end{equation}
for any vector field $\bar{X}$ on $\Lambda^m_2E$. Here, $\Omega_2$ is the multisymplectic $(m+1)$-form on $\Lambda^m_2E$ presented in (\ref{can-forms}). In (\ref{HDW1}), we can choose $\bar{X}$ as vertical vector fields with respect to the projection $\bar{\tau}$ as well, \cite{EcheLeon}. 

\textbf{Reduced HDW equations.} An alternative realization of the Hamiltonian dynamics on the present geometry is available in terms of Hamiltonian sections. This time consider the fiber bundle $(\mathsf  J^1\pi^*,\tau,M)$ where the projection is defined to be $\tau=\pi\circ\nu$. A critical point of a Hamiltonian section $h$ is a section $\phi$ of $\tau$ satisfying the reduced HDW equations 
\begin{equation} \label{HDW2}
\phi^*\iota_{X}\Omega_h=0,
\end{equation}
for any vector field $X$ on $\mathsf J^1\pi^*$. In (\ref{HDW2}), we can choose vertical vector fields $X$ with respect to the projection $\tau$, \cite{EcheLeon}. Here, $\Omega_h$ is a $(m+1)$-form on $\mathsf  J^1\pi^*$ defined by
\begin{equation} \label{Omega_h}
\Omega_h=h^*\Omega_2.
\end{equation} 
\noindent
The Liouville form $\theta_h$ is related to $\Omega_h$ though $\Omega_\theta=-d_{\theta}\theta_h$.
Note also that $h^*d\mathcal H$ is identically zero, so, we can rewrite (\ref{Omega_h}) by adding the term $h^*d\mathcal H$ to the right hand side as well. Both of the formulations in (\ref{HDW1})  and (\ref{HDW2}) result with the same dynamics if $\mathcal H$ and $h$ are related as in (\ref{Ham-sec-dens}). Recall the equations given in \eqref{alphaomega} determining the pull-back of the canonical forms, for a section $\zeta$ of the bundle $(\mathsf  J^1\pi^*,\nu,E)$, we have 
\begin{equation} \label{d-commute}
\zeta^*\theta_h=h\circ\eta, \qquad \zeta^*\Omega_h=-d(h\circ\eta).
\end{equation}

\textbf{HDW equations with an Ehresmann Connection.}
One can also write an infinitesimal counterpart of HDW equations using an Ehresmann connection. Recall the fiber bundle $(TJ^1\pi^*,\tau,M)$, see the diagram in \eqref{fibrations}. An Ehresmann connection for this fibration consists on determining a distribution $\mathbf H$ in $J^1\pi^*$ that is complementary to the vertical subbundle $\sV\tau$ with respect to the fibration $\tau$. So that we can write 
\begin{equation}\label{horizontdistrib}
TJ^1\pi^*=\mathbf H\oplus \sV\tau.
\end{equation}
This reads a horizontal projector $\mathbf h$ which maps a tangent vector in $TJ^1\pi^*$ to its horizontal part in $\mathbf H$.

Notice that $\mathbf h$ can be considered as an element of the tensor product $\Lambda^mM\otimes TJ^1\pi^*$ as well. We may perform tensor contraction $\iota_{\bf h}\Omega_h$ where $\Omega_h$ is the multisymplectic $(m+1)$-form defined in \eqref{Omega_h}. This operation results in another $(m+1)$-form. 
If the projector $\bf{h}$ of an Ehresmann connection satisfies
 \begin{equation} \label{HDW3}
\iota_{\bf h}\Omega_h=(m-1)\Omega_h 
    \end{equation}

 An integral section of $\mathbf h$ is a section $\phi$ of the projection $\tau$ so that the image of the tangent mapping of $\phi$ takes values in $\mathbf H$. In other words, if $\phi$ is an integral section of $\mathbf{h}$ then for a projectable vector field $Z$ on $\mathsf  J^1\pi^*$, the following identity holds
\begin{equation}
T\phi\circ T\tau \circ Z = \mathbf{h}\circ Z.
\end{equation}

Any integral section $\phi$ of the connection is a solution of the Hamilton equations. It is a matter of a direct calculation now to show that such an integral section coincides with those solving the HDW equations (\ref{HDW2}). 

\textbf{Local coordinates.} Recall the coordinates $(x^i,u^\alpha)$ on the total manifold $E$, and the the coordinates $(x^i,u^\alpha,p, p^i_\alpha)$ on $\Lambda^m_2E$. In terms of the coordinates, the relationship (\ref{Ham-sec-dens}) between a Hamiltonian section and Hamiltonian density is computed to be 
\begin{equation} \label{h-incoord}
 h(x^i,u^\alpha,p^i_\alpha)=(x^i, u^\alpha, p=-H(x^i,u^\alpha,p^i_\alpha), p^i_\alpha).
  \end{equation}
Accordingly, the Hamiltonian density can be written as    
\begin{equation} \label{HamFunc}
\mathcal H(x^i,u^\alpha, p, p^i_\alpha)=(p+H(x^i,u^\alpha,p^i_\alpha))d_mx 
  \end{equation}
where the local function $H$ in this definition is called the Hamiltonian function. We remark that the Hamiltonian function $H$ in (\ref{HamFunc}) must not be confused with the globally defined mapping $\bar H$ satisfying $\mathcal H=\bar H\eta$. In this local picture, the $(m+1)$-form $\Omega_h$ on $\mathsf  J^1\pi^*$ defined in (\ref{Omega_h}) becomes
  \begin{equation} 
  \Omega_h=\d H\wedge\d_mx-\d p{^i_\alpha}\wedge\d u{^\alpha}\wedge(\partial_i\lrcorner\d_mx).
    \end{equation}
In local coordinates, the Ehresmann connection has the form
\begin{equation} \label{h}
\mathbf{h}=dx^j\otimes \left(\frac{\partial }{\partial x^j}+\Gamma^\alpha_j\frac{\partial }{\partial u^\alpha}+
\Gamma^i_{\alpha j}\frac{\partial }{\partial p^i_\alpha}\right),
\end{equation}
where $\Gamma^u_j$ and $\Gamma^i_{\alpha j}$ are Christoffel symbols.  A section $\phi$ of $\tau$ can be written in coordinates as $(x^i,\phi^\alpha(x),\phi^i_\alpha(x))$. A section is an integral section of the connection $\mathbf{h}$ if 
\begin{equation}
\Gamma^\alpha_i=\frac{\partial\phi^\alpha}{\partial x^i},\qquad \Gamma^i_{\alpha j}=\frac{\partial\phi^i_\alpha}{\partial x^j}.
\end{equation}
Straightforward computations show that locally, the HDW equations in (\ref{HDW2}) and (\ref{HDW3}) are the same as the system of equations
\begin{equation} \label{locHDW}
\frac{\partial\phi^\alpha}{\partial x^i}=\frac{\partial H}{\partial p^i_\alpha}\circ\phi ,\quad  \frac{\partial\phi^i_\alpha}{\partial x^i}=-\frac{\partial H}{\partial u^\alpha}\circ\phi. 
\end{equation}

\subsection{A Hamilton-Jacobi theory for HDW Equations} \label{HJ-Sec}

Consider a bundle $(E,\pi,M)$, and recall the fibrations exhibited in diagram (\ref{fibrations}). Let us now choose a connection $\mathbf{h}$ on the bundle $(\mathsf  J^1\pi^*,\tau,M)$ determining the horizontal subbundle $\mathbf{H}$. 

Assume a section $\bar{\gamma}$ of the fibration $(\Lambda^m_2E,\bar{\nu},E)$, and referring to this section, we will define a connection on $(E,\pi,M)$ as follows. First, define a section $\gamma$ of $(\mathsf  J^1\pi^*,\nu,E)$ by taking the composition of $\bar{\gamma}$ and the projection $\mu$, that is  $\gamma=\mu\circ \bar{\gamma}$. Now, take a vector field $X$ on $E$ and choose a vector field $Y$ on $\mathsf  J^1\pi^*$ projected to $X$. Then, a connection on $\pi$ is defined to be 
\begin{equation} \label{red-con}
\mathbf{h}^\gamma:T_eE\longrightarrow \mathbf{H}_e^\gamma, \qquad X(e)\mapsto T\nu\circ \mathbf{h}\circ Y\circ \gamma(e).
\end{equation}
Here, $\mathbf{H}^\gamma$ is the horizontal bundle complementing $\sV\pi$ in a direct sum decomposition of $TE$.  
A direct observation tells us that the definition in (\ref{red-con}) results in the same connection if one changes the chosen field $Y$ with another vector field projected on $X$. Here is a commutative diagram summarizing the present situation: 
\begin{equation}
\xymatrix{ \mathsf  J^1\pi^* \ar[rr]^Y 
\ar[dd]^{\nu} && T(\mathsf  J^1\pi^*)\ar[dd]^{T\nu} \ar[r]^{\quad \mathbf{h}}  & \mathbf{H}\ar[dd]^{T\nu} 
  \\
  \\
E \ar@/^1pc/[uu]^{\gamma}\ar[rr]^X  &&TE \ar[r]^{\quad \mathbf{h}^\gamma} & \mathbf{H}^\gamma
}
\end{equation}
Let us state now a Hamilton-Jacobi theorem for the present setting. For this we consider a connection $\mathbf{h}$ of the bundle $(\mathsf J^1\pi^*,\tau,M)$ satisfying the HDW equations (\ref{HDW3}).
\begin{theorem} \label{HJ-thm}
Assume that $\bar{\gamma}$ is a closed section of $(\Lambda^m_2E,\bar{\nu},E)$, and that the section $\gamma=\mu\circ \bar{\gamma}$ induces a flat connection $\bf{h}^\gamma$ referring to the definition \eqref{red-con}. Then the following conditions are equivalent:
\begin{enumerate}
\item If $\sigma$ is an integral section of $\bf h^\gamma$ then $\gamma\circ\sigma$ is a an integral section of $\bf h$,
\item $ h\circ\gamma$ is closed, where $h$ is the Hamiltonian section.
\end{enumerate} 
\end{theorem} \label{HJ-multi}
In order to prove this theorem in a coordinate invariant way, we first present two lemmas.  Accordingly, we recall the diagram (\ref{fibrations}) and identify in it the sections that we recently introduced.
 \begin{equation} \label{fibrations-2}
\xymatrix{ \Lambda^m_2E \ar[rr]_\mu 
\ar[dr]^{\bar{\nu}} && \mathsf J^1\pi^*   \ar[dl]_{\nu} \ar@/_1pc/[ll]_{h} 
  \\
&E \ar@/^1pc/[ul]^{\bar{\gamma}} \ar@/_1pc/[ur]_{\gamma}    \ar[d]^\pi
 \\
& M \ar@/^1pc/[u]^{\sigma}  
}
\end{equation}
Notice that the bundle projections allow the definition in $\sT( \mathsf  J^1\pi^*)$ of two vertical subbundles namely $\sV\tau$ and $\sV\nu$ being the kernels of the tangent mappings $\sT\tau$ and $\sT\nu$, respectively.

\begin{lemma} \label{Lemma-1}
Let $\eta$ be a section of the fibration $\nu$. On the image space of $\eta$ we have
\begin{equation}\label{vectorfield}
\sV\tau=\sT\eta( \sT E)\oplus_{\mathsf J ^1\pi^*}\sV\nu.
\end{equation}
\end{lemma}
\textbf{Proof of Lemma \ref{Lemma-1}.}
First notice that $\sT\eta$ is injective, and that $\dim\sV\tau$ equals to the sum of $\dim T\eta(TE)$ and $\dim\sV\nu$. Assume an element $w$ in the intersection $\sT\eta(TE)\cap\sV\nu$. So that, there exists a vector $v$ in $TE$ so that $w=\sT\eta(v)$. On the other hand, being a vertical vector we have $\sT\nu(w)=0$. Accordingly,  
\begin{equation}
\sT\nu\circ \sT\eta(v)=\sT (\nu\circ\eta)(v)=id_{ \sT E}(v) =0
\end{equation} 
which implies that $v=0$, and it follows that $w=\sT\eta(v)=0$. $\blacksquare$

A corollary of Lemma \ref{Lemma-1} is that on the image space of $\eta$, each vector field $Y$ on $\mathsf J^1\pi^*$ with values in $\sV\tau$ can be written as
\begin{equation}\label{vectorfield}
Y=\eta_*(X)+Z 
\end{equation}
where $X$ is a vector field on $E$ and $Z$ is a vector field on $\mathsf J^1\pi^*$ that takes values in $\sV\nu$. 

\begin{lemma} \label{Lemma-2}
Let $\sigma$ be an integral section of the connection $\bf h^\gamma$ in \eqref{red-con} which is the reduced form of the connection $\mathbf{h}$ solving the HDW equations (\ref{HDW3}), and consider a section $\eta$ of the projection $\nu$. Then, recalling $\Omega_h$ in \eqref{Omega_h}, 
\begin{equation}
({\eta\circ\sigma})^*\iota_{Z}\Omega_h=0
\end{equation}
for any vector field $Z:\mathsf J^1\pi^*\to\sV\nu$.
\end{lemma}
\textbf{Proof of Lemma \ref{Lemma-2}.}
In coordinates, we have $ {\eta\circ\sigma}(x)=\Big(x^i,\sigma^\alpha(x),\eta^i_\alpha(\sigma^\beta(x)) \Big).$
Since $\sigma$ is an integral section of $\bf h^\gamma$ we have
\begin{equation}\label{lemma2sigma}
\frac{\partial\sigma^\alpha}{\partial x^i}=\frac{\partial H}{\partial p^i_\alpha}.
\end{equation}
A straightforward calculation shows that for any vector field $ Z=Z^i_\alpha\frac{\partial}{\partial p^i_\alpha}$
we have 
$$ ({\eta\circ\sigma})^*\iota_{Z}\Omega_h= Z^i_\alpha\Big( \frac{\partial H}{\partial p^i_\alpha}-\frac{\partial\sigma^\alpha}{\partial x^i}\Big) \d_mx. $$
Therefore from (\ref{lemma2sigma}) we obtain the lemma. $\blacksquare$

Now we are ready to prove Theorem \eqref{HJ-multi}. 

\textbf{Proof of Theorem \ref{HJ-multi}.} 
$(1)\Longrightarrow(2)$: Assume now that $\sigma$ is an integral section of $\bf h^\gamma$ and that $\gamma\circ\sigma$ is a solution of the HDW equations, that is, $\gamma\circ\sigma$ is an integral section of the connection $\bf h$. We want to show that the form
$$ h\circ \gamma:E\to \Lambda^m_2E   $$
is closed. For an arbitrary set $\{w_1,\dots,w_{m+1}\}$ of vector fields on the base manifold $M$, we introduce the following notation $(\gamma\circ\sigma)_* w_i=T(\gamma\circ\sigma)w_i=\hat{T}w_i$ and $(h\circ\gamma\circ\sigma)_*w_i=T(h\circ\gamma\circ\sigma)w_i=\tilde{T}w_i$, so we compute 
\begin{align*}
(\gamma\circ\sigma)^*(\iota_X\Omega_h)(w_1,\dots,w_{m+1})&=(\iota_X\Omega_h)\Big((\hat{T} w_1,\dots, \hat{T} w_{m+1}\Big)\\
&=\Omega_h\Big(X,\hat{T} w_1,\dots, \hat{T} w_{m+1}\Big)\\
&= h^*\Omega_2\Big(X,\hat{T}w_1,\dots, \hat{T}w_{m+1}\Big) \\
&=\Omega_2\Big(h_*X,\tilde{T}w_1,\dots, \tilde{T}w_{m+1}\Big)=0
\end{align*}
which is valid for any vector field $X$. In particular, we choose $X$ in the form $\gamma_*S$ where $S$ is a vector field on $E$. Then we continue the previous calculation as follows
\begin{align*}
\Omega_2\Big(h_*X,\tilde{T}w_1,\dots, \tilde{T}w_{m+1}\Big)&=\Omega_2\Big((h\circ \gamma)_*S,\tilde{T}w_1,\dots, \tilde{T}w_{m+1}\Big) \\
& = (h\circ\gamma)^*\Omega_2\Big(S,\sigma_*w_1,\dots, \sigma_*w_{m+1}\Big)
\\
& =-d(h\circ\gamma)\Big(S,\sigma_*w_1,\dots, \sigma_*w_{m+1}\Big)  
\end{align*}
where we have employed the second identity in (\ref{d-commute}) in the fourth line of this computation.  
The condition that $\gamma$ is a closed form guarantees that $d(h \circ\gamma)$ is still a two-horizontal form which means that contracted with two vertical vectors of $\sV E$ gives zero. Therefore, we obtain
$$ d(h\circ\gamma)=0.   $$

$(2)\Longrightarrow(1)$: Let us assume now that the second condition in the statement of the theorem holds i.e. $d(h\circ\gamma)=0$ and that $\sigma$ is an integral section of $\bf h^\gamma$. In this case, we want to show that $\gamma\circ\sigma$ is a solution of the HDW equations. For any vector field $S$ on $M$ and for an arbitrary set $\{w_1,\dots,w_{m+1}\}$ of vector fields on $M$, we have
\begin{align*}
 d(h \circ\gamma)\Big(S, \sigma_*w_1,\dots, \sigma_*w_{m+1}\Big)&= -(h \circ\gamma)^*\Omega_2\Big(S, \sigma_*w_1,\dots, \sigma_*w_{m+1}\Big)\\
&=-\gamma^*\Omega_h\Big(S, \sigma_*w_1,\dots, \sigma_*w_{m+1}\Big)\\
&=-\gamma^*\iota_X\Omega_h\Big(\sigma_*w_1,\dots, \sigma_*w_{m+1}\Big)=0
\end{align*}
where $X$ is assumed to be $\gamma_*(S)$. Here, in the second equality we have employed (\ref{alphaomega}). Next we have
$$\gamma^*\iota_X\Omega_h\Big(\sigma_*w_1,\dots, \sigma_*w_{m+1})\Big)=(\gamma\circ\sigma)^*\iota_X\Omega_h\Big(\sigma_*w_1,\dots, \sigma_*w_{m+1}\Big).   $$
Therefore we obtain
$$ (\gamma\circ\sigma)^*\iota_X\Omega_h=0, $$
for any vector field $X=\gamma_*(S)$. What is more, from Lemma \ref{Lemma-2} we know that the above equation is satisfied by any vertical vector field $Z$. This gives that $\gamma\circ\sigma$ is a solution of the HDW equations. $\blacksquare$

\textbf{Local coordinates.} 
Let us assume now that $\lambda=\d S$ where
$$S=S^i(x^i,u^\alpha_j)\partial_i\lrcorner\d_mx$$ 
is a $1$-semibasic form on $E$. Then in local coordinates the equation $\d(h\circ\mu\circ\gamma)=0$ is equivalent to
\begin{equation}\label{redHJ}
\frac{\partial S^i}{\partial x^i }+H\left(x^i, u^\alpha_j,\frac{\partial S_i}{\partial u^\alpha_j}\right)=f(x^i).
\end{equation}
where $f$ is a real valued function on $M$.

\section{Locally Conformal Multisymplectic Framework} \label{Sec-Lcmf}

\subsection{Locally Conformal Multisymplectic Manifolds} \label{Sec-lcms}

In this subsection we are introducing the notion of locally conformal multisymplectic manifold both in global and local pictures. In \ref{HD-lcm-sm} we will see the need for this introduction.  

\textbf{The global definition.} Let $P$ be differentiable manifold equipped with an $r$-form $\Omega_\theta$. We say that the form $\Omega_\theta$ is locally conformal multisymplectic (l.c.m-s.) $r$-form if it is $1$-nondegenerate and if there exists a closed one-form $\theta$, let us name it as conformal Lee-form inspired by  \cite{Lee,vaisman}, on $P$ satisfying
\begin{equation}\label{prove1}
d\Omega_\theta=\theta\wedge\Omega_\theta. 
\end{equation}
It is easy to depict equation \eqref{prove1}. See that
\begin{equation}
d\Omega_\theta=d(\Omega+\theta\wedge \Theta)=d(\theta\wedge \Theta)
\end{equation}
because $d\Omega=0$, recall that $\Omega$ is a multisymplectic form, i.e., it is closed.
Hence,
\begin{equation*}
d\Omega_\theta=d\theta\wedge \Theta-\theta\wedge d\Theta=\theta\wedge \Theta
\end{equation*}
by definition $d\Theta=-\Omega$ and $d\theta=0$ because it is a closed one form. Notice that if we substitute $\Omega$ by 
$\Omega_\theta$, we
maintain the same expression, since
\begin{equation}
\theta\wedge \Omega=\theta\wedge (\Omega+\theta\wedge \Theta)
\end{equation}
because the triple wedge product $\theta\wedge \theta\wedge \Theta=0$.
So, we can replace $\theta\wedge \Omega$ by $\theta\wedge \Omega_\theta$ in this case and we have \eqref{prove1}.

The triple $(P,\Omega_\theta, \theta)$ is a locally conformal multisymplectic  (l.c.m-s.) manifold, whereas the pair of differential forms $(\Omega_\theta, \theta)$, is a l.c.m-s. structure. If $\theta$ is exact we say that the triple is a  globally conformal multisymplectic  (g.c.m-s.) manifold. 

\textbf{The local definition.} 
In a local picture, a locally conformal multisymplectic structure may be encoded in an open covering $\{U_A\}$ of $P$, and a set of real valued functions $\sigma_A$ on each chart so that the local $r$-form $e^{-\sigma_A}\Omega_\theta\vert_A$ is closed. Here, $\Omega_\theta\vert_A$ denotes the restriction of the l.c.m-s. form $\Omega_\theta$  to the open set $U_A$. Accordingly, we define local $k$-forms on each $U_A$ by
\begin{equation} \label{rel-omega}
\Omega_A:=e^{-\sigma_A}\Omega_\theta\vert_A.
\end{equation}
Notice that, these local forms are closed and $1$-nondegenerate. That is, each pair $(U_A,\Omega_A)$ is a (local) multisymplectic manifold. The definition in \eqref{rel-omega} also determines that the restriction $\theta\vert_A$ of the conformal Lee-form to the open neighbourhood $U_A$ is precisely equal to the exterior derivative $d\sigma_A$ of the local function $\sigma_A$. On the other hand, if we glue the local multisymplectic $r$-forms $\Omega_A$, they result in a (real) line bundle $L\mapsto P$ valued $r$-form $\tilde{\Omega}$ on $P$. Eventually, we have two global forms on $P$, namely a real valued $r$-form $\Omega_\theta$ and a line bundle valued $r$-form $\tilde{\Omega}$. We denote the local realizations of these forms as $\Omega_\theta\vert_A$ and $\Omega_A$, respectively. Further, these local pictures are related through \eqref{rel-omega}. This observation lead us to the local definition of l.c.m-s. manifolds. A $1$-nondegenerate $r$-form $\Omega$ on a manifold $P$ is a l.c.m-s. if there exists a family of local functions $\sigma_A$ on each element of a local covering $U_A$ so that the local $r$-forms $e^{-\sigma_A}\Omega$ are closed. In accordance with this, a manifold $P$ admitting such a $r$-form $\Omega$ is called a l.c.m-s. manifold.  We have already shown that the global definition implies the local definition. But further, we can argue that, if one starts with the local definition, then arrives at the global definition after a gluing process.    

\textbf{Submanifolds.} We have exhibited some distinguished submanifolds of the multisymplectic manifolds in Subsection \ref{Sec-MM}, namely $l$-isotropic submanifolds, $l$-coisotropic submanifolds and $l$-Lagrangian submanifolds. The definitions of these submanifolds rely on $l$-orthogonality in (\ref{l-ortho}). So that, we can easily argue that these definitions are also valid for l.c.m-s. category as well. But in this case, we substitute the role of the multisymplectic form $\Omega$ with a l.c.m-s form $\Omega_\theta$. In the following subsection we shall be determining $l$-Lagrangian submanifolds more concretely on some particular cases of  l.c.m-s. manifolds.

\subsection{Examples of Locally Conformal Multisymplectic Manifolds}

Here we present some immediate examples of l.c.m-s. manifolds.

\textbf{L.c.m-s. structure on the bundle of $k$-forms.}
We present an example of l.c.m-s. manifold in terms of the bundle $\Lambda^kN$ of $k$-forms on a manifold $N$. Recall the canonical forms $\Theta$ and $\Omega$ on $\Lambda^kN$ given in \eqref{Theta} and \eqref{alphaomega}, respectively. $\Lambda^kN$ admits a l.c.m-s. structure. Actually, there is a nice characterization of this canonical l.c.m-s. structure. To this end, we consider a closed one-form $\vartheta$ on $N$ then, by pulling this form back to $\Lambda^kN$ via $\zeta:\Lambda^kN\mapsto N$, define a closed one-form $\theta=\zeta^*\vartheta$. We define the Lichnerowicz differential using this one form by 
\begin{equation}
d_\theta \zeta:=d\zeta +\theta\wedge \zeta
\end{equation}
where $d$ denotes the exterior (deRham) derivative. Here, $\zeta$ is a differential form. Notice that $d_\theta$ is a differential
operator of order $1$. We refer \cite{GuLi84, HaRy99} for more details on the Lichnerowicz differential. For the present purpose, we determine a $k+1$-form on  $\Lambda^kN$ as follows 
\begin{equation} 
\Omega_\theta:=-d_\theta\Theta=-d\Theta+\theta\wedge\Theta =\Omega+\theta\wedge\Theta,
\end{equation} 
Now, it is a matter of a direct calculation to show that the triple $(\Lambda^kN,\Omega_{\theta},\theta)$ is a l.c.m-s. manifold with the conformal Lee-form $\theta$. 

\textbf{L.c.m-s. structure on the space of two-horizontal forms.} Let  
$(E,\pi, M)$ be a fiber bundle over the base $M$. Recall the space of two-horizontal $m$-forms $(\Lambda^m_2E,\bar{\tau},M)$ defined in (\ref{multimom}). The total space $\Lambda^m_2E$ is equipped with the canonical $m$-form $\Theta_2$, and the multisymplectic $(m+1)$-form $\Omega_2$ with local characterizations  (\ref{can-forms}). We refer once more the commutative diagram given in (\ref{fibrations}). Now, take a closed one-form $\vartheta$ on the base manifold $M$, then pull this form back to $\Lambda^m_2E$ by means of the projection $\bar{\tau}$. We denote this one-form by $\theta=\bar{\tau}^*\vartheta$. Then, by employing the Lichnerowicz differential $d_\theta$,
define an $(m+1)$-form 
\begin{equation} \label{Omega_2t}
\Omega_{2,\theta}:=-d_\theta \Theta_2.
\end{equation}
Notice that $\Omega_{2,\theta}$ is a l.c.m-s. structure on $\Lambda^m_2E$ with the Lee-form $\theta$. We denote this l.c.m-s. manifold as 
\begin{equation} \label{2-hor-theta}
\Lambda^m_2E_\theta:=(\Lambda^m_2E,\Omega_{2,\theta},\theta).
\end{equation}

There exists a nice interpretation of Lagrangian submanifolds of l.c.m-s. manifolds in this particular case. For this, referring to the diagram (\ref{fibrations-2}), consider a section $\bar\gamma$ of the fibration $\hat{\nu}$ of  $\Lambda^{m}_2E$. Let us pull the Lee-form $\theta$ by means of this section. So that we compute 
\begin{equation}
\bar\gamma^*\theta=\bar\gamma^*(\pi\circ\nu)^*\vartheta=(\pi\circ\nu\circ\bar\gamma)^*\vartheta=\vartheta.
\end{equation}
Now we pull the l.c.m-s. $(k+1)$-form $\Omega_{2,\theta}$ by $\bar\gamma$, this gives
\begin{equation}\label{alphaomegaconformal}
\begin{split}
\bar\gamma^*\Omega_{2,\theta}&=-\bar\gamma^*d_\theta\Theta_2=-\bar\gamma^*(d\Theta_2-\theta\wedge\Theta_2)=-(d\bar\gamma-\bar\gamma^*\theta\wedge\bar\gamma^*\Theta_2)\\ &=
-d\bar\gamma+\vartheta\wedge\bar\gamma=-d_\vartheta\bar\gamma. 
\end{split}
\end{equation}
where we have employed the identities in (\ref{alphaomega}) in the fourth equality, and the Lichnerowicz differential $d_\vartheta:=d-\vartheta\wedge$ in the last equality. An immediate conclusion from (\ref{alphaomegaconformal}) is that a form $\bar\gamma$ is closed with respect to the Lichnerowicz differential $d_\vartheta$ if and only if $\bar\gamma^*\Omega_{2,\theta}$ vanishes identically. Therefore, we obtain that the image of $\bar\gamma$ is a Lagrangian submanifold of $\Lambda_\theta^kE$ if and only if it is closed with respect to the Lichnerowicz differential $d_\vartheta$.

\subsection{Hamiltonian Dynamics on Locally Conformal Multisymplectic Manifolds} \label{HD-lcm-sm}

In this subsection, we present the gluing problem of local HDW equations and give an answer to it. These discussions will also exhibit the motivation of the introduction of l.c.m-s. To this end, consider a fiber bundle $(E,\pi,M)$ and the multi-momentum bundle $\Lambda^m_2E$ which is canonically multisymplectic. Let us start with a set $\{U_A\}$ of open charts for the base manifold $M$, and then determine the family $\{V_A\}$ of open sets covering $\Lambda^m_2E$. Here, a local chart $V_A$ is defined to be the pre-image $\bar{\tau}^{-1}(U_A)$ of $U_A$. This induces a local projection $\bar{\tau}_A$ from $V_A$ to $U_A$ as well. On each chart $V_A$, assume local multisymplectic forms $\Omega_A$, and define local Hamiltonian sections $\mathcal H_A$ from $V_A$ to $\Lambda^{m+1}U_A$. A critical point of $\mathcal H_A$ is a section $\phi_A$ of the projection $\bar{\tau}$ that satisfies
\begin{equation} \label{l-HDW-Eq}
\phi_A^*\iota_{X_A}(\Omega_A+d\mathcal H_A)=0,
\end{equation}
for any vector field $X_A$ on $V_A$. Equation (\ref{l-HDW-Eq})is the HDW equation given in (\ref{HDW1}) but in the local chart $V_A$. As we glue these pieces local dynamics in order to arrive at a global picture, we need to impose the invariance of the dynamical equations. This implies that the local sections $\phi_A$'s are global. Then the gluing problem consists on determining which is the most general global geometry admitting such a local realization. In the conformal case, this result is more general than in global multisymplectic manifolds. Actually, after performing such an analysis one arrives at the locally conformal multisymplectic category introduced in Subsection \ref{Sec-lcms}. Let us depict this in more detail in terms of the dynamical equations.

\textbf{HDW equations on l.c.m-s. manifolds.} Consider two local dynamics in the non-trivial intersection $V_A\cap V_B$ of two charts. We search here a passage between the two local dynamics. For this, we start by multiplying \eqref{l-HDW-Eq} by a scalar $\lambda_{BA}$. This manipulation preserves the section $\phi_A$, that is $\phi_A=\phi_B$, in the coordinate change if both the multisymplectic form $\Omega_A$ and the exterior derivative $d\mathcal H_A$ of the Hamiltonian section are modified by $\lambda_{BA}\Omega_A$ and $\lambda_{BA}d\mathcal H_A$, respectively. Notice that, the scalars satisfy the cocyle property 
\begin{equation} \label{cocyle}
\lambda_{CA}=\lambda_{CB}\lambda_{BA}
\end{equation}
in the nontrivial intersection of three charts $V_A$, $V_B$ and $V_C$. So that, the scalars can be written as $\lambda_{BA}=e^{\sigma_A-\sigma_B}$ in terms of the local functions $\sigma_A$'s and $\sigma_B$'s. Substitution of this into \eqref{l-HDW-Eq} reads that
\begin{equation}
\phi_A^*\iota_{X_A}(e^{\sigma_A}\Omega_A+e^{\sigma_A}d\mathcal H_A)=0.
\end{equation}
We will examine the differential forms in this equation one by one.  
The family of forms $e^{\sigma_A}\Omega_A$ on each $V_A$ defines a l.c.m-s. structure $(\Omega_\theta,\theta)$ where the local picture of the differential forms are $\Omega_\theta\vert_A=e^{\sigma_A}\Omega_A$ and $\theta\vert_A=d\sigma_A$, respectively. Let us notice that 
\begin{equation}
e^{\sigma_A}d\mathcal H_A=d(e^{\sigma_A}\mathcal H_A)-e^{\sigma_A}\theta\mathcal H_A= d_\theta(e^{\sigma_A}\mathcal H_A) 
\end{equation}
where we have employed the Lichnerowicz differential $d_\theta$. In the light of the cocycle property (\ref{cocyle}), the set of local Hamiltonian sections $\{\mathcal H_A\}$, obeying the identification $e^{\sigma_B}\mathcal H_B=e^{\sigma_A}\mathcal H_A$, determine a global Hamiltonian section $\tilde {\mathcal H}$. Notice that $\tilde {\mathcal H}$ is a section from the multimomentum bundle $\Lambda^m_2 E$ to the total space of the line bundle $L\mapsto \Lambda^m M$, and locally $\tilde{\mathcal{H}}\vert_A=\mathcal H_A$. On the other hand the local sections $\{e^{\sigma_A}\mathcal H_A\}$ give a global Hamiltonian section $\mathcal H$ in \eqref{Ham-sec-dens}. To sum up, we have two global sections $\tilde {\mathcal H}$ and $\mathcal H$ with local realizations, $\mathcal H_A$ and $e^{\sigma_A}\mathcal H_A$, respectively. 
Eventually, by collecting all these observations, we arrive at the locally conformal generalization of the (extended) HDW equations as follows
\begin{equation}
\phi^*\iota_X(\Omega_{2,\theta}+d_\theta\mathcal H)=0,
\end{equation}
where $\Omega_{2,\theta}$ is l.c.m-s. $(m+1)$-form in (\ref{Omega_2t}).

\textbf{Reduced HDW equation on l.c.m-s. manifolds.} It is possible to carry this formulation to the reduced HDW on $\mathsf J^1\pi^*$. Referring to the relationship (\ref{Ham-sec-dens}) between Hamiltonian sections and Hamiltonian densities, we define an $(m+1)$-form on $\mathsf J^1\pi^*$  by
\begin{equation} \label{Omega-theta-h}
(\Omega_\theta)_h:=h^*(\Omega_{2,\theta}+d_\theta\mathcal H)=h^*\Omega_{2,\theta}
\end{equation}
The Lee form for $(\Omega_\theta)_h$ is $\theta_h$. This is easy to see, since
\begin{equation*}
(\Omega_\theta)_h=h^*\Omega_\theta=h^*\Omega+h^*\theta\wedge h^*\Theta,
\end{equation*}
this reads
\begin{equation*}
(\Omega_\theta)_h=\Omega_h+\theta_h\wedge \Theta_h, \ \text{that is, by definition equal to} \ -d_\theta\Theta_h.
\end{equation*}
So, $\theta_h$ is a Lee form for $(\Omega_\theta)_h$.

A critical point of $h$ is a section $\phi:M\to \mathsf J^1\pi^*$ which is a solution of the reduced HDW equations if
\begin{equation}\label{conformalHDW2}
\phi^*\iota_X(\Omega_\theta)_h=0,
\end{equation}
for any vector field $X$ on $\mathsf J^1\pi^*$. If the local picture of the Lee-form is $\theta=\theta_idx^i$ and, locally, $\phi(x)=(x^i,\partial\phi^\alpha(x),\partial\phi^i_\alpha(x))$ then the locally conformal HDW equation (\ref{conformalHDW2}) can be written as
\begin{equation}\label{conformalHDW2coordinates}
\frac{\partial\phi^\alpha}{\partial x^i}=\frac{\partial H}{\partial p^i_\alpha}\circ\phi ,\quad  \frac{\partial\phi^i_\alpha}{\partial x^i}=-\frac{\partial H}{\partial u^\alpha}\circ\phi+\theta_i\phi^i_\alpha . 
\end{equation}
\begin{lemma}
The equation 
\begin{equation}
\phi^*\iota_X(\Omega_\theta)_h=0
\end{equation}
holds for any vector field $X$ on $\mathsf J^*\pi$ if and only if it holds for any vector field $Y: \mathsf J^1\pi^* \to\sV\tau$.
\end{lemma}


\textbf{HDW equation in terms of a connection.} In Subsection \ref{HDW-Sec}, we have presented the HDW equation in three different formulations, whereas the Hamilton-Jacobi formalism in Subsection \ref{HJ-Sec} was given through a Ehresmann connection. Accordingly, we now present the locally conformal HDW equation in terms of an Ehresmann connection and in the following subsection we shall be writing the corresponding HJ formalism. For this, we start by introducing an Ehresmann connection for the bundle $\mathsf J^1\pi^*\to M$ and we employ a horizontal distribution ${\bf H}$ satisfying (\ref{horizontdistrib}). Let $\bf h$ be the associated horizontal projector for ${\bf H}$ mapping a tangent vector in $T\mathsf J^1\pi^*$ to its horizontal part in $\mathbf H$.

\begin{theorem} \label{lcHDWcon}
If the horizontal projector $\bf h$ of an Ehresmann connection satisfies  
\begin{equation} \label{lcHDWcon-eqn}
i_{\bf h}(\Omega_\theta)_h=(m-1)(\Omega_\theta)_h 
\end{equation}
than each integral section of $\bf h$ is a solution of the locally conformal HDW equation \eqref{conformalHDW2}.
\end{theorem}

\textbf{Proof of Theorem \ref{lcHDWcon}.}
The proof of the above theorem is just a matter of a direct calculation. We present this in terms of the coordinates. 
Recall the local picture of the horizontal projector  $\bf h$ in \eqref{h}. In these coordinates, the $(m+1)$-form $(\Omega_\theta)_h$ determined in \eqref{Omega-theta-h} turns out to be
\begin{equation}
(\Omega_\theta)_h=\Big( \frac{\partial H}{\partial u^\alpha}-\theta_ip_\alpha^i\Big)du^\alpha\wedge\d_mx+\frac{\partial H}{\partial p_i^\alpha}dp_i^\alpha\wedge\d_mx  \d H\wedge\d_mx-\d p{^i_\alpha}\wedge\d u{^\alpha}\wedge(\partial_i\lrcorner\d_mx).
\end{equation}
Then contraction of this form with the projector \eqref{h} is computed to be  
\begin{equation}
\begin{split}
\iota_{\bf h}(\Omega_\theta)_h= &\Big(m\frac{\partial H}{\partial u^\alpha}+ \Gamma_{\alpha j}^j-m\theta_ip_\alpha^i \Big)du^\alpha\wedge d_mx+
\Big(m\frac{\partial H}{\partial p^i_\alpha}-\Gamma_i^\alpha \Big)dp^i_\alpha\wedge d_mx \\ &-
(m-1)dp^i_\alpha\wedge du^\alpha\wedge (\partial_i\lrcorner\d_mx)
\end{split}
\end{equation}
Notice that the equation in (\ref{lcHDWcon-eqn}) is exactly the local equations 
\begin{equation}
\Gamma_i^\alpha=\frac{\partial H}{\partial p_i^\alpha}, \qquad \Gamma_{\alpha j}^j=-\frac{\partial H}{\partial u^\alpha}+\theta_ip_\alpha^i.
\end{equation} 
$\blacksquare$

\subsection{A Hamilton-Jacobi Theory for L.c.m-s. Hamiltonian Dynamics}

We have presented the Hamilton-Jacobi theorem \ref{HJ-thm} for the HDW equations in Subsection \ref{HJ-Sec}. In this section, we aim at to generalize this theory to the Hamiltonian dynamics on l.c.m-s.  manifolds. Consider a connection  ${\bf h}$ in the bundle $\mathsf J^1\pi^*$ and, by employing a section $\bar\gamma$, define a connection ${\bf h}^\gamma$  in the same way as in (\ref{red-con}). For a more detailed discussion of a Hamilton-Jacobi theorem on a locally conformal symplectic manifolds one can see \cite{EsLeSaZa19}.  Here is a generalization of Lemma \ref{Lemma-2}. 
\begin{lemma}
If $\sigma$ is an integral section of $\bf h^\gamma$ and $\eta:E\to \mathsf J^1\pi^*$ a section. Then 
$$({\eta\circ\sigma})^*\iota_{Z}(\Omega_\theta)_h=0$$
for any vector field $Z:\mathsf J^1\pi^*\to\sV\nu$.
\end{lemma}

We will present now a generalization of a Hamilton-Jacobi theorem to a locally conformal multisymplectic setting.

\begin{theorem} \label{HJ-lcHDW}
Assume that $\bar{\gamma}$ is a closed section and that the section $\gamma=\mu\circ \bar{\gamma}$ induces a flat connection $\bf{h}^\gamma$ on $E\to M$. Then the following conditions are equivalent: \begin{enumerate}
\item If $\sigma$ is an integral section of $\bf h^\gamma$, then $\gamma\circ\sigma$ is a solution of the conformal Hamilton equations  
$$ \phi^*(\iota_X(\Omega_\theta)_h)=0  $$
\item The $(m+1)$-form $ h\circ\gamma$ is closed with respect to Lichnerowicz differential
$$ d_\vartheta( h\circ\gamma)=0.  $$
\end{enumerate}
\end{theorem}

\textbf{Proof of Theorem \ref{HJ-lcHDW}.}
$(1)\Longrightarrow(2)$: Let us assume now that $\sigma$ is an integral section of $\bf h^\gamma$ and $\gamma\circ\sigma$ is a solution of (\ref{conformalHDW2}). First of all, if the connection $\bf h^\gamma$ is flat we can consider an integral section $\sigma$ of it. The condition that $\gamma\circ\sigma$ is a solution of Hamilton equations is equivalent to stating that $\mu\circ\gamma\circ\sigma$ is an integral section of the connection $\bf h$. In particular we have
$$  (\gamma\circ\sigma)^*(\iota_X(\Omega_\theta)_h)=0 $$
for any vector field $X$ on $\mathsf J^1\pi^*$. We have to show now that the form $ h\circ\gamma:E\to \Lambda^m_2E$ is closed. For any $(w_1,..., w_{m+1})$ tangent vectors on $M$ we have, for $\hat{T}w_i=T(\gamma\circ\sigma)w_i$, and $\tilde{T}w_i=T(h\circ\gamma\circ\sigma)w_i,$

\begin{align*}
(\gamma\circ\sigma)^*(\iota_X(\Omega_\theta)_h)(w_1,...,w_{m+1})&=(\iota_X(\Omega_\theta)_h)\Big(\hat{T}w_1,..., \hat{T}w_{m+1}\Big)\\
&=(\Omega_\theta)_h\Big(X,\hat{T}w_1,..., \hat{T}w_{m+1}\Big)\\
&= \Omega_{2,\theta}\Big(h_*X, \tilde{T}w_1,..., \tilde{T}w_{m+1}\Big)
\end{align*} 
\noindent
that is equal to zero.
\noindent
The above equation is satisfied by any vector field $X$ on $\mathsf J^1\pi^*$. In particular we can consider a vector field of the form $X=\gamma_*(S)$ where $S$ is any vector field (vertical or not) on $E$. Then we have  
\begin{align*}
\Omega_{2,\theta} \Big(h_*X,\tilde{T}w_1,..., \tilde{T}w_{m+1}\Big)=&(h\circ\gamma)^*\Omega_\theta\Big(S,T\sigma(w_1),..., T\sigma(w_{m+1})\Big)\\
=&-d_\vartheta(h\circ\gamma)\Big(S,T\sigma(w_1),..., T\sigma(w_{m+1})\Big)
\end{align*}      
The condition that $\gamma$ is a closed form guarantees that $d_\vartheta(h\circ\gamma)$ is still a two-horizontal form which means that contracted with two vertical vectors gives zero. Therefore, we obtain
$$ d_\vartheta(h\circ\gamma)=0.   $$

$(2)\Longrightarrow(1)$: Assume now that (2) holds i.e. $d_\vartheta(h\circ\gamma)=0$ and that $\sigma$ is an integral section of $\bf h^\gamma$. We have to show that $\gamma\circ\sigma$ is a solution of the Hamilton equations. For any vector field $Y$ on $M$ and $(w_1,..., w_{m+1})$ being tangent vectors on $M$ we have
\begin{align*}
d_\vartheta(h\circ\gamma)\Big(S, T\sigma(w_1),..., T\sigma(w_{m+1})\Big)&= -(h\circ\gamma)^*\Omega_{2,\theta}\Big(S, T\sigma(w_1),..., T\sigma(w_{m+1})\Big)\\
&=-(\gamma)^*(\Omega_\theta)_h\Big(S, T\sigma(w_1),..., T\sigma(w_{m+1})\Big)\\
&=-(\gamma)^*\iota_X(\Omega_\theta)_h\Big(T\sigma(w_1),..., T\sigma(w_{m+1})\Big)=0
\end{align*}
where $X=\gamma_*S$ and in the second equality we used (\ref{alphaomega}). Next we have
$$(\gamma)^*\iota_X(\Omega_\theta)_h\Big(T\sigma(w_1),..., T\sigma(w_{m+1})\Big)=(\gamma\circ\sigma)^*\iota_X(\Omega_\theta)_h\Big(w_1,..., w_{m+1}\Big).   $$
Therefore we obtain
$$ (\gamma\circ\sigma)^*\iota_X(\Omega_\theta)_h=0, $$
for any vector field $X=\gamma_*S$. What is more, from Lemma 1 we know that the above equation is satisfied by any vector field $Z: \mathsf J^1\pi^*\to\sV(\pi\circ\bar\nu)$. Therefore, it implies that $\gamma\circ\sigma$ is a solution of the Hamilton equations. $\blacksquare$

\section{Examples}

\subsection{Conformal time-dependent mechanics}

An example to illustrate our theory is the gluing problem of time-dependent systems. In this case, we consider the product manifold $E=\mathbb R\times M$ with a trivial fibration over the base space $M$. We choose $t$ as the real variable in $\mathbb{R}$, and the local coordinates $(t,u^\alpha)$ on $E$. Any section $ \sigma$ is $\sigma(t)=(t,\sigma^\alpha(t)).$
We introduce coordinates $(t,u^\alpha,u_t^\alpha)$ in $\mathsf J^1\pi$, $(t, u^\alpha, p,p^t_\alpha)$ in the multimomentum bundle $\Lambda^m_2 E$, hence $(t, u^\alpha, p^t_\alpha)$ in $\mathsf J^1\pi^*$.  From now on we will skip index $t$ in $p^t_\alpha$ for the sake of the simplicity of the notation.
The first jet of $\phi$ in coordinates reads $(t,\phi^\alpha, \phi^\alpha_t)$. We consider a system described by the Hamiltonian section
$$ h(t,u^\alpha,p_\alpha)= \Big( t, u^\alpha, -\frac{1}{2}\delta ^{\alpha \beta }p_\alpha p_\beta ,  p_\alpha  \Big) $$
where $H(t,u^\alpha,p_\alpha)=\frac{1}{2}\delta ^{\alpha \beta }p_\alpha p_\beta$ is a Hamiltonian function. The Hamiltonian density is therefore given by
$$ \mathcal H(t,u^\alpha,p_\alpha)=p+H(t,u^\alpha,p_\alpha)= \Big(p+\frac{1}{2}\delta ^{\alpha \beta }p_\alpha p_\beta \Big)dt$$
The Liouville and multisymplectic forms are 
\begin{equation}
\Theta_2=p\d t+p_\alpha\d u^\alpha, \qquad \Omega_2=-\d p\wedge\d t-\d p_\alpha\wedge\d u^\alpha. 
\end{equation}
Let us take a closed one form $\vartheta=\vartheta_t dt$. We have $\theta=(\pi\circ\nu)^*\vartheta$. In coordinates $ \theta=\theta_t dt$, where $\vartheta_t=\theta_t$. The condition that $\theta$ is closed is fulfilled automatically. The locally conformal multisymplectic form is
$$\Omega_{2,\theta}=-d_\theta\Theta_2=\Omega_2+\theta\wedge\Theta_2=-\d p\wedge\d t-\d p_\alpha\wedge\d u^\alpha+ \theta_ip_\alpha\d u^\alpha\wedge\d t.$$
The equation $\phi^*\iota_X(\Omega_\theta)_h=0$ results in the system  $$\frac{\partial\sigma^\alpha}{\partial t}=p_\alpha
,\qquad  \frac{\partial p_\alpha}{\partial t}=\theta_tp_\alpha .$$
We move now to the HJ problem. We start with a closed section  
$$\bar\gamma(t,u^\alpha)= \rho(t,u^\alpha)\d t+\gamma_\alpha(t,u^\alpha)\d u^\alpha. $$
The assumption of the theorem is that $\bar\gamma$ is closed, which implies that
$$ d\bar\gamma= \frac{\partial\rho}{\partial u^\alpha}du^\alpha\wedge\d t - \frac{\partial\gamma_\alpha}{\partial t}\d u^\alpha\wedge\d t+ \frac{\partial\gamma_\alpha}{\partial u^\beta}\d u^\beta\wedge\d u^\alpha=0$$
and this is equivalent to
$$  \frac{\partial\rho}{\partial u^\alpha} - \frac{\partial\gamma_\alpha}{\partial t}=0, \qquad  \frac{\partial\gamma_\alpha}{\partial u^\beta}-\frac{\partial\gamma_\beta}{\partial u^\alpha}=0. $$
The condition $i)$ says that $\sigma$ is an integral section of $\bf h^\gamma$ which, in coordinates, reads
$$\frac{\partial\sigma^\alpha}{\partial t}=p_\alpha. $$
On the other hand, if a section
$$\mu\circ\bar\gamma\circ\sigma:M\to J^1\pi^*, \quad  \mu\circ\bar\gamma\circ\sigma(x^i)= \gamma_\alpha(t,\sigma^\alpha(t))\d u^\alpha   $$
is a solution of the locally conformal HDW equations (\ref{conformalHDW2coordinates}) then
\begin{equation}\label{exampleequations1}
\frac{\partial\sigma^\alpha}{\partial t}=p_\alpha ,\quad  \frac{\partial \gamma_\alpha}{\partial t}= -p_\beta\frac{\partial\gamma_\beta}{\partial u_\alpha} +\theta_t\gamma_\alpha .
\end{equation}
To see condition $ii)$, the section $h\circ\mu\circ\gamma$ is
$$h\circ\mu\circ\bar\gamma(t,u^\alpha)=h\Big(t,u^\alpha, \gamma_\alpha(t,u^\alpha)\Big)=-H(t,u^\alpha, p_\alpha)\d t+\gamma_\alpha(t,u^\alpha)\d u^\alpha, $$
and the Lichnerowicz differential gives 
$$ d_\vartheta( h\circ\mu\circ\bar\gamma)=\Big( -\frac{\partial H}{\partial u^\alpha} - \frac{\partial H}{\partial p_\beta}\frac{\partial\gamma_\beta}{\partial u_\alpha}   - \frac{\partial\gamma_\alpha}{\partial t} +\vartheta_t\gamma_\alpha \Big)\d u^\alpha\wedge\d t +\frac{\partial\gamma_\alpha}{\partial u^\beta}\d u^\beta\wedge\d u^\alpha=0.  $$
Therefore we obtain that condition $ii)$ in coordinates is rewritten as
\begin{equation}
- p_\beta\frac{\partial\gamma_\beta}{\partial u_\alpha}  -\frac{\partial \gamma_\alpha}{\partial t} +\vartheta_t(t)\gamma_\alpha =0, \qquad 
\frac{\partial\gamma_\alpha}{\partial u^\beta}-\frac{\partial\gamma_\beta}{\partial u^\alpha}=0.
\end{equation}
Let us notice that the second equation is fulfilled under the assumption that $\gamma$ is a closed section while the first one is the second equation in (\ref{exampleequations1}). It is easy to see then that both of the conditions from HJ theory are indeed equivalent. 
\subsection{Conformal $N$-scalar field}

Another example of our theory is the case of a conformal generalization of a theory of $N$ free scalar fields. A set of $N$ scalar fields $\phi$ is a section of a trivial bundle $E=M\times\mathbb R^N\to M$ i.e., $  \sigma:M\to M\times\mathbb R^N, \ \text{such that}\ \sigma(x^i)=(x^i,\sigma^\alpha(x^i)).$
We introduce coordinates $(x^i,u^\alpha,u^\alpha_i)$ in $\mathsf J^1\pi$, $(x^i, u^\alpha, p,p^i_\alpha)$ in $\Lambda^m_2E$, and $(x^i, u^\alpha, p^i_\alpha)$ in $\mathsf J^1\pi^*$.
The first jet of $\phi$ in coordinates reads $(x^i,\sigma^\alpha, \sigma^\alpha_i)$. We assume that the manifold $M$ is equipped with a metric tensor $g$ which defines a volume form $\sqrt{\det|g|}d_mx$ on $M$. The Lagrangian density of a scalar field theory is a map
$$\mathcal L:\sJ \pi \to\Lambda^mM, \quad  \mathcal L(x^i,u^\alpha,u^\alpha_i)=\frac{1}{2}g^{ij}u^\alpha_iu^\alpha_j\sqrt{\det|g|}d_mx $$
where $L(x^i,u^\alpha,u^\alpha_i)=\frac{1}{2}g^{ij}u^\alpha_iu^\alpha_j\sqrt{\det|g|}$ is a Lagrangian function, i.e., $\mathcal{L}=Ld_mx$. The hamiltonian section associated with $L$ reads
$$ h(x^i,u^\alpha,p^i_\alpha)= \Big( x^i, u^\alpha, -\frac{1}{2}\frac{1}{\sqrt{\det|g|}}g_{ij}p^i_\alpha p^j_\alpha ,  p^i_\alpha  \Big) $$
where $H(x^i,u^\alpha,p^i_\alpha)=\frac{1}{2}g_{ij}p^i_\alpha p^j_\alpha(1/{\sqrt{\det|g|}})$ is the corresponding  Hamiltonian function. The Hamiltonian density is therefore given by
$$ \mathcal H(x^i,u^\alpha,p^i_\alpha)=p+H(x^i,u^\alpha,p^i_\alpha)= \Big(p+\frac{1}{2}g_{ij}p^i_\alpha p^j_\alpha\frac{1}{\sqrt{\det|g|}} \Big)d_mx$$
The Liouville and multisymplectic forms look exactly the same as the ones in \eqref{can-forms}.
Let us take a closed one form $\vartheta=\vartheta_i\d x^i$ on $M$. We have $\theta=(\pi\circ\nu)^*\vartheta$. In coordinates, $ \theta=\theta_idx^i,$ where $\vartheta_i=\theta_i$ and the condition that $\theta$ is closed in coordinates is equivalent to ${\partial\theta_i}/{\partial x^j}={\partial\theta_j}/{\partial x^i}$. The locally conformal multisymplectic form is
$$\Omega_{2,\theta}=-d_\theta\Theta_2=\Omega_2+\theta\wedge\Theta_2=-\d p\wedge\d_mx-\d p_\alpha^i\wedge\d u^\alpha\wedge(\partial_i\lrcorner\d_mx)+ \theta_ip_\alpha^i\d u^\alpha\wedge\d_mx.$$
The locally conformal HDW equation \eqref{conformalHDW2} turns out to be
\begin{equation}
\frac{\partial\sigma^\alpha}{\partial x^i}=\frac{1}{\sqrt{\det|g|}} g_{ij}p^j_\alpha
,\quad  \frac{\partial p^i_\alpha}{\partial x^i}=\theta_ip^i_\alpha . 
\end{equation}
Let us move now to the Hamilton-Jacobi problem. We start with a closed section 
$$ \bar\gamma(x^i,u^\alpha)= \rho(x^i,u^\alpha)\d_mx+\gamma^i_\alpha(x^i,u^\alpha)\d u^\alpha\wedge(\partial_i\lrcorner\d_mx). $$
The assumption of the theorem is that $\bar \gamma$ is closed which means that
$$ d\bar\gamma= \frac{\partial\rho}{\partial u^\alpha}du^\alpha\wedge\d_mx - \frac{\partial\gamma^i_\alpha}{\partial x^i}\d u^\alpha\wedge\d_mx+ \frac{\partial\gamma^i_\alpha}{\partial u^\beta}\d u^\beta\wedge\d u^\alpha\wedge(\partial_i\lrcorner\d_mx)=0.$$
This gives that 
$$  \frac{\partial\rho}{\partial u^\alpha} - \frac{\partial\gamma^i_\alpha}{\partial x^i}=0, \qquad  \frac{\partial\gamma^i_\alpha}{\partial u^\beta}-\frac{\partial\gamma^i_\beta}{\partial u^\alpha}=0. $$
The condition $i)$ says that $\sigma$ is an integral section of $\bf h^\gamma$ which, in coordinates, becomes
$$\frac{\partial\sigma^\alpha}{\partial x^i}=\frac{1}{\sqrt{\det|g|}} g_{ij}p^j_\alpha. $$
On the other hand, if a section
$$\mu\circ\bar\gamma\circ\sigma:M\to \mathsf J^1\pi^*, \quad  \mu\circ\bar\gamma\circ\sigma(x^i)= \gamma^i_\alpha(x^i,\sigma^\alpha(x^i))\d u^\alpha\wedge(\partial_i\lrcorner\d_mx)    $$
is a solution of the conformal Hamilton equations (\ref{conformalHDW2coordinates}) then
\begin{equation}\label{exampleequations}
\frac{\partial\sigma^\alpha}{\partial x^i}=\frac{1}{\sqrt{\det|g|}} g_{ij}p^j_\alpha ,\quad  \frac{\partial \gamma^i_\alpha}{\partial x^i}= -\frac{1}{\sqrt{\det|g|}} g_{ij}p^j_\beta\frac{\partial\gamma^i_\beta}{\partial u_\alpha}  +\theta_i\gamma^i_\alpha .
\end{equation}
Let us write the second condition in HJ theorem. In coordinates, the section $h\circ\mu\circ\bar\gamma$ is
\begin{equation}
\begin{split}
h\circ\mu\circ\bar\gamma(x^i,u^\alpha)&=h\Big(x^i,u^\alpha, \gamma^i_\alpha(x^i,u^\alpha)\Big)\\&=-H(x^i,u^\alpha, \gamma^i_\alpha(x^i,u^\alpha)) \d_mx+\gamma^i_\alpha(x^i,u)\d u^\alpha\wedge(\partial_i\lrcorner\d_mx)
\end{split}
\end{equation}
and the Lichnerowicz differential of it reads
\begin{equation}
\begin{split}
d_\vartheta( h\circ\mu\circ\bar\gamma)= &\Big(  -\frac{\partial H}{\partial u^\alpha} - \frac{\partial H}{\partial p^i_\beta}\frac{\partial\gamma^i_\beta}{\partial u_\alpha} - \frac{\partial\gamma^i}{\partial x^i} +\theta_i\gamma^i_\alpha \Big)\d u^\alpha\wedge\d_mx \\ & +\frac{\partial\gamma^i_\alpha}{\partial u^\beta}\d u^\beta\wedge\d u^\alpha\wedge(\partial_i\lrcorner\d_mx)=0.  
\end{split}
\end{equation}
Therefore, the Hamilton--Jacobi equation reads:
\begin{equation}
\frac{1}{\sqrt{\det|g|}} g_{ij}p^j_\beta\frac{\partial\gamma^i_\beta}{\partial u_\alpha}  + \frac{\partial \gamma^i_\alpha}{\partial x^i} -\vartheta_i(x^j)\gamma^i_\alpha =0, 
 \qquad 
\frac{\partial\gamma^i_\alpha}{\partial u^\beta}-\frac{\partial\gamma^i_\beta}{\partial u^\alpha}=0.
\end{equation}
For $N=1$ one obtains a theory of a free scalar field.

\section{Cauchy data space on a locally conformal multisymplectic manifold.}

The Cauchy data space allows us to relate the finite-dimensional and the infinite-dimensional formulation of field theory. It was shown in \cite{CaLeDiVa15} that the Hamilton-Jacobi theory for multisymplectic manifolds can be formulated in the Cauchy space, so here we show an extension of this construction to the locally conformal multisymplectic framework.

\subsection{A space of Cauchy data}

\textbf{Cauchy surface.} Assume that there is a slicing of the base manifold $M$ into the Cartesian product of $\mathbb R$ and a compact, oriented and embedded submanifold $\Sigma$, called Cauchy surface \cite{LeonDiegoMerino,Sni84}. We determine this by the following diffeomorphism
\begin{equation} \label{slicing}
\chi_M:\mathbb R\times \Sigma\to M, \qquad (t,y)\longmapsto\chi_M(t,y).
\end{equation}
Here, we consider that $t$ is a global coordinate chart for $\mathbb R$. Define a vector field $\xi_M$ on $M$, that is an infinitesimal generator of the diffeomorphism $\chi_M$, by pushing the vector field ${\partial}/{\partial t}$ on $\mathbb{R}$ by the slicing mapping \eqref{slicing}, that is 
\begin{equation} \label{inf-gen-slicing}
\xi_M=(\chi_M)_*\frac{\partial}{\partial t}\in \mathfrak{X}(M).
\end{equation}
The slicing diffeomorphism  manifests that the dimension of $M$ is equal to $m=n+1$ where $n$ is the dimension of $\Sigma$. We further assume the existence of a volume form $\eta_\Sigma$ on $\Sigma$ decomposing the volume form $\eta$ on $M$ as a product $dt\wedge \eta_\Sigma$, and that the area of the Cauchy surface is the unit under the measure $\eta_\Sigma$. 
As it can easily be predicted the existence of such geometries is not guaranteed for an arbitrary volume manifold. Nevertheless, we omit a detailed discussion of topological properties that ensure this, and assume that any functional analytic obstruction is automatically satisfied.

For each fixed $t$ in $\mathbb R$ the slicing \eqref{slicing}, we define an embedding $(\chi_M)_t$ of $\Sigma$ in $M$. 
Assume also that there exists $t_0$ in $\mathbb R$ such that the image space $(\chi_M)_{t_0}(\Sigma)$ is precisely $\Sigma$. We denote all possible embeddings of $\Sigma$ into the product manifold $M$ by $\widetilde{M}$. Notice that, $\widetilde{M}$ is diffeomorphic to  $\mathbb{R}$. We are interested in studying these embeddings, the so-called $\chi$-sections, of the Cauchy surface into the total spaces of the bundles exhibited in the right hand side of diagram \eqref{fibrations}.

\textbf{Manifolds of embeddings, $\wdt E$ and $\wdt{\mathsf J^1\pi^*}$.}
We define the space of the projection $\pi$ as follows
\begin{equation} \label{wE}
\wdt E:=\{\sigma_\Sigma(t):\Sigma\to E \ \vert \ \exists t\in  \mathbb{R}\text{ such that }\pi\circ\sigma_\Sigma(t)=(\chi_M)_t   \}.  
\end{equation}
It is obvious that $\wdt E$ is not a finite dimensional manifold, since it is a manifold of mappings (embeddings) \cite{KrMi97,Mi80}. 
Similarly to \eqref{wE}, we define the manifold $\wdt{\mathsf J^1\pi^*}$ of embeddings of $\Sigma$ as the space of sections $\phi$ of the fibration $\tau:\mathsf J^1\pi^*\mapsto M$ projecting to $(\chi_M)_t$ for some $t$ that is
\begin{equation} \label{wJ*}
\wdt{\mathsf J^1\pi^*}:=\{\phi_\Sigma(t):\Sigma\to \mathsf J^1\pi^*  \ \vert \ \exists t\in  \mathbb{R}\text{ such that } \tau \circ\phi_\Sigma(t)=(\chi_M)_t   \}. 
\end{equation} 
The following diagram summarizes the definitions (\ref{wE}) and (\ref{wJ*}) done for the elements of the manifolds of embeddings $\wdt E$ and $\mathsf J^1\pi^*$. Referring to the right hand side of diagram (\ref{fibrations-2}), we plot  
 \begin{equation} \label{fibrations-4}
\xymatrix{&&& \mathsf J^1\pi^*   \ar[d]^{\nu} \ar@/^2pc/[dd]^{\tau}
  \\
&&& E \ar@/^1pc/[u]_{\gamma} \ar[d]^{\pi}   
\\
\Sigma \ar@{.>}[urrr]_{\sigma_\Sigma(t)} \ar@{.>}[uurrr]^{\phi_\Sigma(t)}   \ar@{.>}[rrr]_{(\chi_M)_t} &&& M
\ar@/^1pc/[u]_{\sigma}  \ar@/^2pc/[uu]^{\phi}
 \\
}
\end{equation} 
It is immediate to observe from this diagram that, from a section $\sigma$ of the fibration $\pi$, one can define an element $\sigma_\Sigma (t)$ of $\wdt E$ by a composition $\sigma \circ (\chi_M)_t$. Additionally, a section $\phi$ of $\tau$ induces an element $\phi_\Sigma(t)$ by $\phi \circ (\chi_M)_t$. The inverses of these statements are also true. That is, elements of $\wdt E$ and $\wdt{\mathsf J^1\pi^*}$ determine sections of $\pi$ and $\tau$, respectively. This is a crucial observation, since it is enabling us to formulate an infinite dimensional framework. 

\textbf{Bundles of $\wdt E$ and $\wdt{\mathsf J^1\pi^*}$.}
We can define a map projecting an embedding $\sigma_\Sigma(t)$ in $\wdt E$ to the real number $t$ satisfying the definition in \eqref{wE}, that is, one has the following fibration
\begin{equation}
\wdt {\pi}:\wdt E \longrightarrow \wdt M\cong \mathbb{R}, \qquad \sigma_\Sigma(t)\mapsto t.
\end{equation}
Since the base manifold $\wdt{M}$ is diffeomorphic to $\mathbb{R}$, a section of $\wdt {\pi}$ is a differentiable curve. We denote a section by $\sigma_\Sigma:\mathbb{R}\to \wdt E$. For a real number $t$, every value $\sigma_\Sigma(t)$ of $\sigma_\Sigma$ is an embedding of $\Sigma$ into $E$ satisfying definition \eqref{wE}. 

By mimicking the procedure done in the case of $\tilde{E}$, we define a bundle structure 
\begin{equation}
\wdt{\tau}:\wdt{\mathsf J^1\pi^*}\mapsto \wdt{M}\cong \mathbb{R}, \qquad \phi_\Sigma(t)\mapsto t.
\end{equation}
As in the previous case, a section of $\wdt{\tau}$ is a curve in $\wdt{\mathsf J^1\pi^*}$. We denote a section by $\phi_\Sigma:\mathbb{R}\mapsto \wdt{\mathsf J^1\pi^*}$, where at each point $t$, $\phi_\Sigma(t)$ turns out to be an embedding of $\Sigma$ into $\mathsf J^1\pi^*$ satisfying the definition \eqref{wJ*}. Notice that, referring to the bundle ${\nu}:{\mathsf J^1\pi^*} \mapsto E$, we can construct a fibration
\begin{equation}
\wdt{\nu}:\wdt{\mathsf J^1\pi^*} \mapsto \wdt E, \qquad \phi_{\Sigma}(t) \mapsto \nu \circ \phi_{\Sigma}(t).
\end{equation}
Starting with a section $\gamma$ of the projection $ {\mathsf J^1\pi^*}\to E$, we define a section $\wdt \gamma$ of $ \wdt{\mathsf J^1\pi^*}\to \wdt E$ through the equation 
\begin{equation} \label{tilde-gamma}
\wdt \gamma:\wdt E\longrightarrow \wdt{\mathsf J^1\pi^*}, \qquad  \sigma_{\Sigma}(t) \mapsto \gamma\circ \sigma_{\Sigma}(t). 
\end{equation}
We show a diagram illustrating the fibrations and the sections. The following diagram and the one in \eqref{fibrations-4} are complementing each other, while fixing finite and infinite dimensional objects.
 \begin{equation} \label{fibrations-5}
\xymatrix{ \wdt{\mathsf J^1\pi^*}   \ar[d]^{\wdt \nu} \ar@/_1pc/[dd]_{\wdt \tau}
  \\
\wdt{E} \ar[d]^{\wdt \pi} \ar@/_2pc/[u]_{\wdt \gamma}     
\\
 \wdt{M}\cong \mathbb{R}
\ar@/_2pc/[u]_{\sigma_\Sigma }  \ar@/^3pc/[uu]^{\phi_\Sigma }
 \\
}
\end{equation}

\textbf{Tangent spaces of $\wdt E$ and $\wdt{\mathsf J^1\pi^*}$.}
One can define tangent vectors and differential forms on $\wdt E$ as follows. A tangent vector $\wdt X$ in $T_{\sigma_\Sigma(t)}\tilde{E}$ is a differentiable map $\Sigma\to\sT E$ in the following commutative diagram.
\begin{equation} \label{materialvf}
\xymatrix{ 
& &\sT \wdt E \ar[rr]^{T\pi} \ar[d] & & \sT M\ar[d]  \\  \Sigma\ar@/_2pc/[rrrr]_{(\chi_M)_t} \ar[rr]_{\sigma_\Sigma(t)} \ar[urr]^{\wdt X}
  && \wdt E \ar@/_1pc/[u]_{X} \ar[rr]^{\pi}   && M \ar@/_1pc/[u]_{\xi_M}
}
\end{equation} 
Let $\alpha$ be a $(k+n)$-form on $E$. Then,  a $k$-form $\wdt\alpha$ on $\wdt E$ is defined to be 
\begin{equation} \label{diff-mm}
\wdt\alpha(\sigma_\Sigma(t))(\wdt X_1,\dots,\wdt X_k)=\int_\Sigma\sigma_\Sigma(t)^*(\iota_{\wdt{X}_1\wedge \dots \wedge \wdt{X}_k}\alpha).
\end{equation}
As a particular instance, we can define a one-form $\wdt\eta$ on $\wdt E$ referring to the volume $(n+1)$-form $\eta$. See that, in this case, $\wdt\eta$  coincides with $\widetilde{\pi}^*dt$. Similarly, an element $\wdt{Y}$ of the tangent space $T_{\phi_{\Sigma}(t)} \wdt{\mathsf J^1\pi^*}$ is a differentiable mapping $\wdt{Y}:\Sigma \mapsto T\mathsf J^1\pi^*$ projecting to $\phi_{\Sigma}(t)$. 

\begin{equation} \label{materialvf}
\xymatrix{ 
& &\sT \wdt{\mathsf J^1\pi^*} \ar[rr]^{T\pi} \ar[d] & & \sT M\ar[d]  \\  \Sigma\ar@/_2pc/[rrrr]_{(\chi_M)_t} \ar[rr]_{\sigma_\Sigma(t)} \ar[urr]^{\wdt X}
  && \wdt{\mathsf J^1\pi^*} \ar@/_1pc/[u]_{X} \ar[rr]^{\pi}   && M \ar@/_1pc/[u]_{\xi_M}
}
\end{equation} 

It is important for the present discussion to note also that the following relationship holds for the exterior derivative \cite{CaLeDiVa15,LeonDiegoMerino}
\begin{equation} \label{alpha}
d\wdt \alpha = \wdt{d\alpha},
\end{equation}  
for any differential form $\alpha$. So, we can define a two-form $\wdt\Omega_h$ on $\wdt{\mathsf J^1\pi^*}$ by employing this definition in the multisymplectic $(m+1)$-form $\Omega_h$, defined in \eqref{Omega_h}. Hence $(\wdt{\mathsf J^1\pi^*},\wdt\Omega_h)$ becomes a presymplectic manifold due to identity \eqref{alpha}.

\textbf{Local realizations.} We choose coordinates $(t,x^1,...,x^n)$ on $M$ such that locally $\Sigma_t$'s are given by the level sets of $t$. For the total spaces, this induces local coordinates $(t,x^i,u^\alpha,p^t_\alpha,p^i_\alpha)$ on $J^1\pi^*$, and $(t,x^i,u^\alpha, p, p^t_\alpha,p^i_\alpha)$ on $\Lambda^{n+1}_2E$. A point in $\wdt E$ is given by specifying functions $u^\alpha(\cdot)$ that depend on the coordinates on $\Sigma_t$, i.e. $(x^i)$, $i=1,...,n$ whereas points of $\wdt{J^1\pi^*}$ and $\wdt{\Lambda^{n+1}_2E}$ are given by specifying functions $p^t_\alpha(\cdot)$, $p^i_\alpha(\cdot)$ and $p(\cdot)$ that depend on $(x^i)$. So, elements of $\wdt E$, $\wdt{J^1\pi^*}$, and $\wdt{\Lambda^{n+1}_2E}$ are respectively given by
\begin{equation}
\begin{split}
&(t,u^\alpha(\cdot))\in \wdt E, \quad t\in\mathbb R, \quad u^\alpha=u^\alpha(x^1,...,x^n),
\\
&(t, u^\alpha(\cdot), p^t_\alpha(\cdot), p^i_\alpha(\cdot)) \in \wdt{J^1\pi^*},
\\
&(t, u^\alpha(\cdot), p(\cdot), p^t_\alpha(\cdot), p^i_\alpha(\cdot)) \in  \wdt{\Lambda^{n+1}_2E}.
\end{split}
\end{equation}

\subsection{Dynamics on the space of Cauchy data and HJ theory} \label{sec-dsCd}

Recall the presymplectic  manifold $(\wdt{\mathsf J^1\pi^*},\wdt\Omega_h)$ defined in the previous subsection. Now, by referring to the identifications given between finite and infinite dimensional frameworks, we write the HDW dynamics in terms of the manifold of embeddings \cite{CaLeDiVa15}. 
\begin{theorem} \label{infdimHamEq}
A section $\phi$ is a solution of HDW equation \eqref{HDW2} if and only if the corresponding mapping $\phi_\Sigma(t)$ verifies 
\begin{equation}
\iota_{\dot{\phi}_\Sigma(t)}\wdt\Omega_h=0.
\end{equation}
\end{theorem}
As it is shown in Subsection \ref{HDW-Sec}, it is possible to recast HDW equations in terms of Ehresmann connections. In the present infinite dimensional version, to refer to an Ehresmann connection \eqref{horizontdistrib}, we employ $\mathbf{h}$ as a horizontal lift operator, see \eqref{h-incoord}, from sections of the tangent bundle $TM$ to sections of the tangent bundle $TJ^1\pi^*$. Accordingly, for an Ehresmann connection $\mathbf{h}$, we define a vector field $\widetilde X^\bh$ on $\widetilde{J^1\pi^*}$ as follows: 
\begin{equation} \label{X^h}
\widetilde X^\bh:\widetilde{J^1\pi^*} \longrightarrow T\widetilde{J^1\pi^*}, \qquad \phi_{\Sigma}(t)\mapsto \widetilde X^\bh(\phi_{\Sigma}(t)),
\end{equation}
where $\widetilde X^\bh(\phi_{\Sigma}(t))$, is an element of $T_{\phi_{\Sigma}(t)}\widetilde{J^1\pi^*}$, and it is a differentiable mapping 
\begin{equation}
\widetilde X^\bh(\phi_{\Sigma}(t)):\Sigma \longrightarrow TJ^1\pi^*,\qquad y \mapsto \mathbf{h}(\xi_M\circ \chi_M(t,y)).
\end{equation}
Here, $\chi_M$ is the slicing operator in \eqref{slicing} whereas $\xi_M$ is the infinitesimal generator in \eqref{inf-gen-slicing}. 
In the same way we can construct a vector field $\widetilde X^{\bh^\gamma}$ on $\widetilde E$ by employing the induced connection $\bh^\gamma$, defined in \eqref{red-con}, on $E$. Accordingly,  
\begin{equation}
\widetilde X^{\bh^\gamma}:\widetilde E \longrightarrow T \widetilde E, \qquad \sigma_{\Sigma}(t)\mapsto \widetilde X^{\bh^\gamma}(\sigma_{\Sigma}(t)),
\end{equation}
where $\widetilde X^{\bh^\gamma}(\sigma_{\Sigma}(t))$, being an element of $T_{\sigma_{\Sigma}(t)}\widetilde E$, is a differentiable mapping 
\begin{equation}
\widetilde X^{\bh^\gamma}(\sigma_{\Sigma}(t)):\Sigma \longrightarrow TE,\qquad y \mapsto \bh^\gamma(\xi_M\circ \chi_M(t,y)).
\end{equation}
It is possible to see from the definition of the reduced connection $\bh^\gamma$ in \eqref{red-con} that by pushing the vector field $\widetilde X^\bh$ forward via the projection $\wdt \nu$ (see diagram (\ref{fibrations-5})), one arrives precisely at the vector field $\widetilde X^{\bh^\gamma}$.  

\textbf{Hamilton-Jacobi problem.} Notice that in the infinite dimensional case there is no notion of a Hamilton-Jacobi equation, and neither of its solution. Therefore, our idea is to extrapolate this definition from the finite-dimensional case using the properties that admit such an extrapolation. Accordingly, recall Hamilton-Jacobi Theorem \ref{HJ-thm} for multisymplectic Hamiltonian systems. Assume that $\gamma$ is a solution of this Hamilton-Jacobi problem. Then, according to the definition in (\ref{tilde-gamma}), we can determine a section $\wdt \gamma$ of the fibration $\widetilde{J^1\pi^*}\mapsto \wdt E$. The following theorem states the properties of such a section \cite{CaLeDiVa15}.
\begin{theorem}
A section $\tilde \gamma$, defined through (\ref{tilde-gamma}) using a solution $\gamma$ of Hamilton-Jacobi problem \ref{HJ-thm}, satisfies the following two conditions:
\begin{enumerate}
\item $\tilde{\gamma}^*\wdt\Omega_h=0$, 
\item $\iota_{\tilde{\gamma}_* \widetilde X^{\bh^\gamma}}\wdt\Omega_h\vert_{\sigma_{\Sigma}}=0$ for all $\sigma_{\Sigma}$.
\end{enumerate}
\end{theorem}

\subsection{Locally conformal dynamics on the space of Cauchy data and HJ theory} 

Here, we are generalizing Hamiltonian dynamics and the associated HJ theory presented in Subsection \ref{sec-dsCd} for the case of locally conformal geometry. Our first observation here is the following.
\begin{lemma} \label{preco-lemma}
Referring to the vector field $\wdt X^\bh$ in \eqref{X^h}, 
\begin{equation}
\iota_{\wdt X^\bh}(\wdt{\Omega_\theta})_h=0, \qquad \iota_{\wdt X^\bh}\widetilde{\eta}=1,
\end{equation}
where $(\wdt{\Omega_\theta})_h$ and $\widetilde{\eta}$ are differential forms on $\mathsf J^1\pi^*$ obtained by the method in \eqref{diff-mm}, from $(\Omega_\theta)_h$ in \eqref{Omega-theta-h} and the pull back volume form $\eta$, respectively. 
\end{lemma}

From this lemma, we conclude that the triple $(\wdt{J^1\pi^*}, (\wdt{\Omega_\theta})_h, \wdt{\eta})$ is almost \textit{precosymplectic} system, because the integrated from $\widetilde{\Omega_h}$ is not closed \cite{IbMaSo92}. This is the main difference from the multisymplectic setting, the rest is the same.

Now, we have a following theorem
\begin{theorem}\label{HJinf} The section $\wdt\gamma$ satisfies:\\
i) $\wdt{\gamma}^*(\wdt{\Omega_\theta})_h=0$ \\
ii) $i_{\sT\wdt\gamma(\sigma_E)(\wdt X^{h^\gamma})}(\wdt{\Omega_\theta})_h=0 \text{ for all } \sigma_E\in\wdt E \text{ which is an integral submanifold of } \bf h^\gamma_{\widetilde \pi(\sigma_E)}. $
\end{theorem}
Due to these properties, we say that $\wdt\gamma$ is a solution of a Hamilton- Jacobi equation in an infinite dimensional setting.

The proof is a straightforward generalization of Proof 5.4 in \cite{CaLeDiVa15}. It is shown in \cite{LeonDiegoMerino} that $\wdt{\gamma}^*(\wdt{\Omega_\theta})_h=\widetilde{\gamma^*(\Omega_\theta)_h}$. This means that for $\gamma$ being a solution of a HJ theorem, we have $\gamma^*(\Omega_{\theta})_h= d_\vartheta( h\circ\mu\circ\gamma)=0$, so that i) holds. 

Proof for ii) will be done in coordinates. Assume that $\gamma$ is given by an expression
$$ \gamma(t,x^i,u^\alpha)=(t,x^i, u^\alpha,\gamma_p(t,x^i,u^\alpha), \gamma_{p^t_\alpha}(t,x^i,u^\alpha), \gamma_{p^i_\alpha}(t,x^i,u^\alpha))  $$
and $u^\alpha\circ\sigma_E=\sigma^\alpha_E$. So
$$  T\wdt\gamma(\wdt X(\sigma_E))= \frac{\partial}{\partial t}+\Gamma^\alpha_0(\wdt\gamma(t,\sigma^\alpha_E(\cdot)))\frac{\partial}{\partial u^\alpha}
+\Big( \frac{\partial\gamma^t_\alpha}{\partial t}+  \frac{\partial\gamma^t_\alpha}{\partial u^\beta}\Gamma^\beta_0(\wdt\gamma(t,\sigma^\alpha_E(\cdot)) \Big)\frac{\partial}{\partial p^t_\alpha}+$$
$$+ \Big( \frac{\partial\gamma^i_\alpha}{\partial t}+  \frac{\partial\gamma^i_\alpha}{\partial u^\beta}\Gamma^\beta_0(\wdt\gamma(t,\sigma^\alpha_E(\cdot)) \Big)\frac{\partial}{\partial p^i_\alpha}  $$
and
$$ i_{\sT\wdt\gamma(\sigma_E)(\wdt X^{h^\gamma})}(\wdt{\Omega_\theta})_h=i_{\sT\wdt\gamma(\sigma_E)(\wdt X^{h^\gamma})}\wdt\Omega_{ h}-i_{\sT\wdt\gamma(\sigma_E)}\wdt{h^*\theta\wedge\Theta}  $$ 
An easy computation shows that for any $\wdt\pi^o_1$-vertical tangent vector $\wdt\xi\in\sT_{\wdt\gamma(\sigma_E)}\wdt{J^1\pi^o}$
$$ \wdt\xi=\xi_{u^\alpha}(\cdot)\frac{\partial}{\partial u^\alpha}+ \xi_{p^t_\alpha}(\cdot)\frac{\partial}{\partial p^t_\alpha}+ \xi_{p^i_\alpha}(\cdot)\frac{\partial}{\partial p^i_\alpha}   $$
by the definition of $(\wdt{\Omega_\theta})_h$, the expression $(\wdt{\Omega_\theta})_h\Big(T\wdt\gamma(\sigma_E)(\wdt X^{{\bf h}^\gamma}), \wdt\xi\Big)$ reads
$$(\wdt{\Omega_\theta})_h\Big(T\wdt\gamma(\sigma_E)(\wdt X^{{\bf h}^\gamma}), \wdt\xi\Big)= 
\int_{\Sigma}   \xi_{u^\alpha}\Big( -\frac{\partial\gamma^i_\alpha}{\partial x^i}-\frac{\partial\gamma^i_\alpha}{\partial u^\beta}\frac{\partial\sigma^\beta_E}{\partial x^i}-\frac{\partial\gamma^0_\alpha}{\partial t}-\frac{\partial\gamma^0_\alpha}{\partial u^\beta}\Gamma^\beta_0-\frac{\partial H}{\partial u^\alpha}+\theta_ip^i_\alpha  \Big)d_mx+$$
$$ +\int_{\Sigma}\xi_{p^i_\alpha}\Big( \frac{\partial\sigma^\alpha_E}{\partial x^i}-\frac{\partial H}{\partial p^i_\alpha}\Big)d_mx+ \int_{\Sigma}\xi_{p^t_\alpha}\Big(-\frac{\partial H}{\partial p^t_\alpha}+\Gamma_0^\alpha\Big)  d_mx $$
The second and the third term vanish because $\sigma_E$ is an integral submanifold of a restricted connection and because $\Gamma_0^\alpha=\frac{\partial H}{\partial p^t_\alpha}$. To show that the first term is equal to zero let us notice that the condition $d_\vartheta(h\circ\mu\circ\gamma)=0$ implies in coordinates.
\begin{equation}\label{Cauchyproof}
 \frac{\partial H}{\partial u^\alpha} + \frac{\partial H}{\partial p^i_\beta}\frac{\partial\gamma^i_\beta}{\partial u_\alpha} + \frac{\partial H}{\partial p^0_\beta}\frac{\partial\gamma^0_\beta}{\partial u_\alpha} + \frac{\partial\gamma^i}{\partial x^i}-\frac{\partial\gamma^0}{\partial t} +\theta_i\gamma^i_\alpha =0  
\end{equation} 
Now using again the fact that  $\frac{\partial\sigma^\beta_E}{\partial x^i} =\frac{\partial H}{\partial p^i_\beta}$ and that $d\gamma=0$ one obtains that the first term is equal to the left-hand side of (\ref{Cauchyproof}).
$\blacksquare$.  

\section{Conclusion and future work}

In this paper we have studied the problem of conformal classical fields in finite and infinite dimensional manifolds. For the finite case, we have used the multisymplectic setting, and for the infinite dimensional case we have introduced the Cauchy data space.
The novelty of this work is the introduction of the concept of locally conformal multisymplectic manifolds, as well as to derive a Hamilton-Jacobi theory to solve the dynamics of conformal fields. This Hamilton-Jacobi theory is developed in the conformal multisymplectic framework and in the Cauchy data space.
These results are applied to two examples: conformal time-dependent mechanics and free conformal scalar fields.
In forthcoming papers we aim at developing a discretization of this theory, starting from a discrete Hamilton-Jacobi theory on a multisymplectic manifold. This result will succeed in its applicability in real physical field theories. For example, you can have a field equation as the heat equation on a continuous space, or you can discretize space having values on the vertices of a graph, with the emergent topology determined by the graph's structure. This is common in engineering simulations of heat equations.

\section{Acknowledgements}
This work has been partially supported by MINECO grants MTM2016-76-072-P and the ICMAT Severo Ochoa Project SEV-2011-0087 and SEV-2015-0554. OE grateful to Prof J. \'Sniatycki for many discussions on the field theory. The research of Marcin Zaj\k{a}c was financially supported by the University’s Integrated Development Programme (ZIP) of University of Warsaw co-financed by the European Social Fund under the Operational Programme Knowledge Education Development 2014 - 2020, action 3.5.

\end{document}